\documentclass[onecolumn,nofootinbib]{revtex4}
\usepackage{amsfonts}
\usepackage{amsmath}
\usepackage{amssymb}

\input{tcilatex}
\begin{document}

\title{Gravitational Waves in Locally Rotationally Symmetric (LRS) Class II
Cosmologies}
\author{Michael Bradley$^{1}$, Mats Forsberg$^{1}$ and Zolt\'{a}n Keresztes$%
^{2,3}$}
\address{$^{1}$\textit{{Department of Physics, Ume{\aa } University, 901 87
Ume\aa , Sweden},}\\
michael.bradley@physics.umu.se, forsberg.mats.a.b@gmail.com,\\
$^{2}$ \textit{{Department of Theoretical Physics, University of Szeged,} }\\
\textit{{Tisza Lajos krt 84-86, Szeged 6720, Hungary}, }\\
zkeresztes@titan.physx.u-szeged.hu\\
$^{3}$ \textit{{Department of Experimental Physics, University of Szeged,} }%
\\
\textit{{D\'{o}m T\'{e}r 9, Szeged 6720, Hungary}}}
\date{\today }

\begin{abstract}
In this work we consider perturbations of homogeneous and hypersurface
orthogonal cosmological backgrounds with local rotational symmetry (LRS),
using a method based on the 1 + 1 + 2 covariant split of spacetime. The
backgrounds, of LRS class II, are characterised by that the vorticity, the
twist of the 2-sheets, and the magnetic part of the Weyl tensor all vanish.
They include the flat Friedmann universe as a special case. The matter
contents of the perturbed spacetimes are given by vorticity-free perfect
fluids, but otherwise the perturbations are arbitrary and describe
gravitational, shear, and density waves. All the perturbation variables can
be given in terms of the time evolution of a set of six harmonic
coefficients. This set decouples into one set of four coefficients with the
density perturbations acting as source terms, and another set of two
coefficients describing damped source-free gravitational waves with odd
parity. We also consider the flat Friedmann universe, which~has been
considered by several others using the 1 + 3 covariant split, as a check of
the isotropic limit. In agreement with earlier results we find a
second-order wavelike equation for the magnetic part of the Weyl tensor
which decouples from the density gradient for the flat Friedmann universes.
Assuming vanishing vector perturbations, including the density gradient, we
find a similar equation for the electric part of the Weyl tensor, which was {%
previously unnoticed}. 
\end{abstract}

\maketitle

\section{Introduction}

In light of the recent success in measurements of gravitational waves \cite%
{Abbott2016}, and the consequent opening of a new observational window, it
is of interest to study the propagation of gravitational waves and their
interactions on different cosmological backgrounds to see, for example, what
the effects of anisotropy and/or inhomogeneities are.

The fluctuations in the cosmic microwave background radiation (CMB), the
large-scale structures and the cosmological redshift are well explained by
the $\Lambda$CDM model  \cite{Komatsu49,Spergel48,WMAP9yr,Planck1,Planck2},
which describes an almost homogeneous and isotropic universe with a
cosmological constant and cold dark matter. However, there are some
deviations between the data and the model. For example, the observed power
spectrum of the CMB seems to differs from the $\Lambda$CDM model for large
angles \cite{Bennett46,Oliveria45,Vielva59,PlanckAnomaly}. Since a large
amount of alternative matter is needed to account for the dark sector, a
wide range of alternative cosmological models have also been investigated to
explore if they can explain the current observations \cite%
{alt1,alt2,alt3,alt4,alt5,alt6,alt7,alt8,alt9,alt10,alt11,alt12}. Also,
present redshift studies do not give very strict bounds on the anisotropy in
the expansion \cite{H1,H2,Dec}, making studies of different types of
perturbations on anisotropic cosmological backgrounds of interest. For
earlier works on this see, for example, \cite%
{Doroschkevich,Perko,Tomita,Gumruk,Pereira,Pitrou}, and for different
perturbative methods see \cite{Lifshitz,Bardeen,Stewart,Hawking,Olson}.

In this paper we will use a method based on the 1 + 3 and 1 + 1 + 2
covariant splits~of\mbox{
spacetime~%
\cite{cov1,cov3,cov5,cov6,cov7,Cargese,Bonometto,perturb2,1+1+2,LRSIIscalar,Schperturb,3+1+1}}
to study perturbations 
on anisotropic backgrounds. In the 1 + 3 split there is a preferred timelike
vector, like the 4-velocity of matter, which is used to split tensors into
timelike scalars and spacelike 3-tensors in a way that keeps covariance of
the tensors. Similarly, a further 1 + 2 split can be made with respect to a
spatial direction. This split is natural when there is a preferred spatial
direction on the background, but can be also be used for isotropic
backgrounds. Hence, the direction is fixed by, for example, choosing one of
the perturbed vectors along it. The gauge problem in relativistic
perturbation theory is here avoided by using covariant objects which vanish
on the background for the perturbed quantities \cite{StewartWalker}.

In an earlier paper \cite{GWKS} we considered perturbations on a Kantowski-Sachs
 background, using the 1 + 1 + 2 covariant split of
spacetime \cite{1+1+2}. The perturbations were vorticity-free and the
perturbed spacetime was considered to be described by a perfect fluid. The
perturbations include density fluctuations, shear waves, and pure
gravitational perturbations travelling with the speed of light at leading
order in the high frequency limit. Moreover, beyond this geometrical optics
limit, anisotropic dispersion relations were indicated. The full dynamics
were found to be given by evolution equations for six harmonic coefficients
which decouple into two sub-systems, one with two components describing pure
gravitational degrees of freedom, and one with the remaining four
coefficients where the density gradient acts as a source term.

In this work we extend the previous 1 + 1 + 2 analysis to a wider class of
locally rotationally symmetric (LRS) backgrounds. LRS symmetry means that
spacetime is invariant under rotations around at least one spatial direction
at every point \cite{LRS,MarklundBradley}. The analysis will cover
vorticity-free perturbations of all LRS spatially homogeneous and
hypersurface orthogonal perfect fluid backgrounds with vanishing magnetic
part of the Weyl tensor, except the hyperbolic and closed Friedmann models,
which together with the flat model have been considered by several authors
(see, for example,
\cite%
{cov1,DunsbyBassettEllis,Challinor,perturb4,Gebbie1,Gebbie2,Marteens,Tsagas}
), using the 1 + 3 split. The considered metrics all belong to LRS class II
in the classification of \cite{LRS}. We find the vorticity-free
perturbations of the homogeneous and hypersurface orthogonal LRS II
backgrounds to behave in an analogous way to those in the previous study on
Kantowski-Sachs backgrounds \cite{GWKS} and a similar harmonic
decomposition can be used. Still, all harmonic coefficients can be expressed
in terms of six coefficients and the evolution equations have the same
structure as before, but the behaviour of the solutions of course varies
according to which backgrounds are taken. Finally, as~a~consistency check,
we consider the flat Friedmann background as the isotropic limit of our 1 +
1 + 2 equations. We find that the magnetic part of the Weyl tensor, which in
the general anisotropic case is partly sourced by the density gradient,
satisfies a source-free second-order damped wave equation. This is in
agreement with earlier results, using the 1 + 3 covariant split of
spacetime. A second-order equation is also obtained for the electric part of
the Weyl tensor for the case of pure tensor perturbations , in contrast to
an earlier study \cite{DunsbyBassettEllis}.

The paper is organized as follows: In Section \ref{sectioncovariant} a short
summary of the 1 + 3 and 1 + 1 + 2 covariant splits of spacetime is given.
The LRS backgrounds are discussed in Section \ref{sectionLRS}. In Section %
\ref{sectionharmonic} the harmonic expansion is described and then the
evolution equations for the harmonic coefficients are given in Section \ref%
{sectionevolution}. The high-frequency limit is considered in Section \ref%
{sectionhighfrequency}. The flat Friedmann case is treated in Section \ref%
{sectionFriedmann}. Conclusions are summarised in Section \ref{conclusion}.

We use the signature convention $(-+++)$ and units where $c=1$ and $8\pi G=1$%
.

\section{The 1 + 3 and 1 + 1 + 2 Covariant Splits of Spacetime}

\label{sectioncovariant}

In this section we give a brief summary of the 1 + 3 and 1 + 1 + 2 covariant
splits of spacetime. For~more details on 1 + 3 split the reader is referred
to \cite{Cargese, cov1} and for 1 + 1 + 2 split to \cite{1+1+2, Schperturb}.
A~ summary of the two can also be found in \cite{GWKS}.

A 1+3 split of spacetime is suitable when there is a preferred timelike
vector $u^{a}$. The projection operator onto the perpendicular 3-space is
given by $h_{a}^{b}=g_{a}^{b}+u_{a}u^{b}$ in terms of the 4-metric $g_{ab}$.
With~the help of $h_{ab}$ vectors and tensors can be covariantly decomposed
into spatial and timelike parts. The covariant time derivative and projected
spatial derivative are given by 
\begin{equation}
\dot{\psi}_{a..b}\equiv u^{c}\nabla _{c}\psi _{a...b}\quad \hbox{and}\quad
D_{c}\psi _{a...b}\equiv h_{c}^{f}h_{a}^{d}...h_{b}^{e}\nabla _{f}\psi
_{d...e}~,
\end{equation}%
respectively. The covariant derivative of the 4-velocity, $u^{a}$, can be
decomposed as 
\begin{equation}
\nabla _{a}u_{b}=-u_{a}A_{b}+D_{a}u_{b}=-u_{a}A_{b}+\frac{1}{3}\theta
h_{ab}+\omega _{ab}+\sigma _{ab}~,
\end{equation}%
where the kinematic quantities of $u^{a}$, acceleration, expansion,
vorticity, and shear are defined by $A_{a}\equiv u^{b}\nabla _{b}u_{a}$, $%
\theta \equiv D_{a}u^{a}$, $\omega _{ab}\equiv D_{[a}u_{b]}$, and $\sigma
_{ab}\equiv D_{<a}u_{b>}$, respectively. Here square brackets $[\;]$ denote
anti-symmetrisation, and angular brackets $<\;>$ denote the symmetric and
trace-free part of a~tensor, i.e. $\psi_{<ab>}\equiv \left(h_{(a}^c h_{b)}^d-%
\frac{1}{3}h_{ab}h^{cd}\right)\psi_{cd}$. These quantities, together with
the Ricci tensor (expressed via the Einstein equations by, for example,
energy density $\mu $ and pressure $p$ for a perfect fluid) and the
electric, $E_{ab}\equiv C_{acbd}u^{c}u^{d}$, and magnetic, $H_{ab}\equiv 
\frac{1}{2}\varepsilon _{ade}C^{de}\!\!_{bc}u^{c}$, parts of the Weyl
tensor, are~then used as independent variables. Here $\varepsilon
_{abc}\equiv \eta _{dabc}u^{d}\equiv 4!\sqrt{-g}\delta _{\lbrack
d}^{0}\delta _{a}^{1}\delta _{b}^{2}\delta _{c]}^{3}u^{d}$ is the
three-dimensional volume~element.

The Ricci identities for $u^a$ and the Bianchi identities then provide
evolution equations in the $u^a$ direction and constraints (see for example 
\cite{Cargese}).

A formalism for a further split (1 + 2) with respect to a preferred spatial
vector $n^{a}$ (with $u^{a}n_{a}=0$) was developed in \cite{Schperturb,
1+1+2}. Projections perpendicular to $n^{a}$ are made with $%
N_{a}^{b}=h_{a}^{b}-n_{a}n^{b}$, and in an~analogous way to above, spatial
vectors and tensors may be decomposed into scalars along $n^{a}$ and
perpendicular two-vectors and symmetric, trace-free two-tensors as $A^{a}=%
\mathcal{A}n^{a}+\mathcal{A}^{a}$ , $\omega ^{a}=\Omega n^{a}+\Omega ^{a}$,
and $\sigma _{ab}=\Sigma (n^{a}n^{b}-\frac{1}{2}N_{ab})+2\Sigma
_{(a}n_{b)}+\Sigma _{ab}$. This occurs similarly for $E_{ab}$ and $H_{ab}$
in terms of $\mathcal{E}$, $\mathcal{E}_{a}$, $\mathcal{E}_{ab}$ and $%
\mathcal{H}$, $\mathcal{H}_{a}$, $\mathcal{H}_{ab}$, respectively.
Derivatives along and perpendicular to $n^{a}$ are
\begin{equation}
\widehat{\psi }_{a...b}\equiv n^{c}D_{c}\psi
_{a...b}=n^{c}h_{c}^{f}h_{a}^{d}...h_{b}^{e}\nabla _{f}\psi _{d...e}\;\;%
\hbox{and}\;\;\delta _{c}\psi _{a...b}\equiv
N_{c}^{f}N_{a}^{d}...N_{b}^{e}D_{f}\psi _{d...e}~,
\end{equation}%
respectively. Similarly to the decomposition of $\nabla _{a}u_{b}$, $%
D_{a}n_{b}$, and $\dot{n}_{a}$ can be decomposed into further kinematic
quantities of $n^{a}$ as 
\begin{equation}
D_{a}n_{b}=n_{a}a_{b}+\frac{1}{2}\phi N_{ab}+\xi \epsilon _{ab}+\zeta
_{ab}\quad \hbox{and}\quad \dot{n}_{a}=\mathcal{A}u_{a}+\alpha _{a}~,
\end{equation}%
where $a_{a}\equiv \hat{n}_{a}$, $\phi \equiv \delta _{a}n^{a}$, $\xi \equiv 
\frac{1}{2}\varepsilon ^{ab}\delta _{a}n_{b}$, $\zeta _{ab}\equiv \delta
_{\{a}n_{b\}}$, $\mathcal{A}\equiv n^{a}A_{a}$, and $\alpha _{a}\equiv
N_{a}^{b}\dot{n}_{b}$. The two-dimensional volume element is given by $%
\varepsilon _{ab}\equiv \varepsilon _{abc}n^{c}$ and curly brackets $\{\;\}$
denote the symmetric and trace-free part of 2-tensors. A bar on vector
indices will denote projection onto the 2-sheets, e.g., $\dot{\psi}_{\bar{a}%
}\equiv N_{a}^{\;b}\dot{\psi}_{b}$.

The Ricci and Bianchi identities are then written as constraints and
evolution and propagation equations in the $u^{a}$ and $n^{a}$ directions,
respectively (see \cite{1+1+2}). For the commutation relations between the
differential operators $\;\dot{}\;$, $\;\widehat{}\;$ and $\delta _{a}$ when
acting on scalars, vectors and tensors, see Appendix \ref{commutation}.

\section{Locally Rotationally Symmetric Spacetimes}

\label{sectionLRS}

A spacetime which at each point is invariant under rotations around at least
one spatial direction is referred to as locally rotationally symmetric, or
LRS for short. The corresponding locally maximally symmetric 2-sheets
perpendicular to the isotropy axis are characterized by the 2D curvature
scalar ${\mathcal{R}}=2{\mathcal{K}}/a_2^2$, where $a_2$ is the radius of
curvature (or alternatively the scale factor) and ${\mathcal{K}}=\pm 1$ or 0
for spheres, pseudo-spheres, or planes, respectively.

The perfect fluid LRS spacetimes can be divided into three classes, I, II
and III \cite{LRS, MarklundBradley}. The~Class I metrics are stationary with
nonzero vorticity and vanishing shear and expansion and hence are of limited
interest as cosmological models. Class II is characterized by the fact that
the magnetic Weyl tensor $H_{ab}$, vorticity $\omega _{ab}$, and 2-sheet
twisting $\xi $ all vanish. In general, spacetimes in this class are both
time- and space-dependent and it contains many physically interesting
solutions like spherically-symmetric perfect fluids, the inhomogeneous
Lemaitre-Tolman-Bondi cosmologies, the homogeneous
Kantowski-Sachs and LRS Bianchi I and III cosmologies, and the
flat and hyperbolic Friedmann models. The metrics in LRS class III are
spatially homogeneous with a nonzero twist of the 2-sheets and have
vanishing vorticity and acceleration, as well as vanishing expansion of the
2-sheets. The only models in this class with vanishing magnetic part of the
Weyl tensor are the closed Friedmann~models.

Perturbations of Kantowski-Sachs universes, which are the
hypersurface orthogonal and homogeneous LRS II models with positive
2-curvature, ${\mathcal{R}}>0$, were studied by us in an earlier paper \cite%
{GWKS}. In this paper we extend the analysis to all hypersurface orthogonal
and homogeneous LRS II models with vanishing expansion of the 2-sheets, 
i.e., $\phi=0$. With this last requirement a~similar harmonic decomposition
as in \cite{GWKS} can be used. Fortunately this only excludes the hyperbolic
Friedmann universes. Since the only solutions in LRS class III with
vanishing magnetic part of the Weyl tensor are the closed Friedmann
universes \cite{Marklund}, it means that our analysis will cover all
homogeneous and hypersurface orthogonal LRS backgrounds with a vanishing
magnetic part of the Weyl tensor except the hyperbolic and closed Friedmann
models, which together with the flat model have been studied elsewhere with
the 1 + 3 covariant split (see e.g.,  \cite%
{cov1,DunsbyBassettEllis,Challinor,perturb4,Gebbie2,Marteens,Tsagas}).

\subsection{LRS Class II}

The perfect fluid LRS Class II spacetimes are characterised by $%
H_{ab}=\xi=\omega_{ab}=0$, see, e.g.,~\cite{MarklundBradley, LRS}. In terms
of the quantities defined in Section \ref{sectioncovariant}, the spacetimes
are given by the following scalars: the~energy density $\mu$, the pressure $%
p $, the electric part of the Weyl tensor ${\mathcal{E}}$, the expansion $%
\Theta$, the shear~$\Sigma$, the acceleration ${\mathcal{A}}$, and the
expansion of the 2-sheets, $\phi$. Alternatively, one of the quantities can
be replaced with the 2-curvature of the 2-sheets 
\begin{equation}
{\mathcal{R}}=\frac{2}{3}\left( \mu +\Lambda \right) -2\mathcal{E}-\frac{1}{2%
}\left( \Sigma -\frac{2\Theta }{3}\right) ^{2}+\frac{\phi^2}{2}\; .
\end{equation}

For a complete local description of the geometry, the frame vectors along
the 4-velocity, $u^a$, and the preferred spatial direction, $n^a$, are
needed to construct all Cartan invariants (see e.g., \cite{BradleyMarklund}%
). In~terms of timelike and spacelike coordinates, $t$ and $z$,
respectively, they are given by 
\begin{equation}
e_0=u=X\partial_t+x\partial_z \; ,\quad e_1=n=Y\partial_t + y\partial_z
\end{equation}
where $X$, $x$, $Y$ and $y$ are functions of $t$ and $z$. For spatially
homogeneous spacetimes, where all invariant objects are functions of a
timelike coordinate solely, we may without loss of generality change the
time coordinate so that $X=1$ and $x=0$. Metrics with 4-velocity $u^a$
orthogonal to the hypersurfaces of homogeneity are obtained by putting $Y=0$%
. The assumption of $Y\neq 0$ implies that ${\mathcal{R}}=0$ (see \cite%
{MarklundBradley}), and give rise to tilted models of Bianchi types V or I.

The quantities $\{\mu,p,{\mathcal{E}},\Theta,\Sigma,{\mathcal{A}}%
,\phi,X,Y,x,y\}$, which describe the spacetime, are subject to integrability
conditions given by commutator equations between $e_0$ and $e_1$, the Ricci
equations for $u^a$ and $n^a$, and some of the Bianchi identities (see \cite%
{MarklundBradley,BradleyKarlhede, BradleyMarklund}). Einstein's equations
are imposed through the Ricci tensor, which for a perfect fluid is given by $%
\mu$, $p$ and $u^a$.

\subsubsection{Homogeneous and Hypersurface Orthogonal LRS II Metrics}

With the assumptions $x=Y=0$, it follows that the acceleration vanishes, ${%
\mathcal{A}}=0$, \cite{MarklundBradley}. The system then reduces to the
following evolution equations: 
\begin{equation}
\dot{\mu}=-\Theta \left( \mu +p\right) \ ,  \label{continuity0}
\end{equation}%
\begin{equation}
\dot{\Theta}=-\frac{\Theta ^{2}}{3}-\frac{3}{2}\Sigma ^{2}-\frac{1}{2}\left(
\mu +3p\right) +\Lambda \ ,  \label{expansiondot0}
\end{equation}%
\begin{equation}
\dot{\Sigma}=-\left( \frac{2}{3}\Theta +\frac{1}{2}\Sigma \right) \Sigma -{%
\mathcal{E~}},
\end{equation}%
\begin{equation}
\dot{\mathcal{E}}=\left( \frac{3}{2}\Sigma -\Theta \right) {\mathcal{E}}-%
\frac{1}{2}(\mu +p)\Sigma ~,
\end{equation}%
\begin{equation}
\dot{\phi}=-\left( \frac{1}{3}\Theta -\frac{1}{2}\Sigma \right) \phi \;,
\end{equation}%
where a dot indicates derivative with respect to $t$, and to the constraints 
\begin{equation}
\phi \Sigma =\phi {\mathcal{E}}=0 ~,
\end{equation}%
\begin{equation}
3\mathcal{E}=-2\left( \mu +\Lambda \right) -3\Sigma ^{2}+\frac{2}{3}\Theta
^{2}+\Sigma \Theta \ -\frac{3}{2}\phi ^{2}\;.  \label{EWeyl0}
\end{equation}

\subsubsection{Homogeneous and Hypersurface Orthogonal LRS II Metrics with $%
\protect\phi=0 $}

\label{sectionphi0} For $\phi=0$ the system reduces to 
\begin{equation}
\dot{\mu}=-\Theta \left( \mu +p\right) \ ,  \label{continuity}
\end{equation}%
\begin{equation}
\dot{\Theta}=-\frac{\Theta ^{2}}{3}-\frac{3}{2}\Sigma ^{2}-\frac{1}{2}\left(
\mu +3p\right) +\Lambda \ ,  \label{expansiondot}
\end{equation}%
\begin{equation}
\dot{\Sigma}=\frac{2}{3}\left( \mu +\Lambda \right) +\frac{\Sigma ^{2}}{2}%
-\Sigma \Theta -\frac{2}{9}\Theta ^{2}\ ,  \label{Sigmadot}
\end{equation}
with ${\mathcal{E}}$ given algebraically by 
\begin{equation}
3\mathcal{E}=-2\left( \mu +\Lambda \right) -3\Sigma ^{2}+\frac{2}{3}\Theta
^{2}+\Sigma \Theta \ .  \label{EWeyl}
\end{equation}

Hence, given an equation of state $p=p\left( \mu \right) $ the solutions are
completely determined in terms of~$\Sigma $, $\Theta $ and $\mu $, and the
nonzero zeroth order quantities are given by the set $S^{(0)}=\{\Sigma,%
\Theta,\mu,p,{\mathcal{E}}\}$.

The 2-curvature is now given by  
\begin{equation}
\mathcal{R}=\frac{2}{3}\left( \mu +\Lambda \right) -2\mathcal{E}-\frac{1}{2}%
\left( \Sigma -\frac{2\Theta }{3}\right) ^{2}=2\left( \mu +\Lambda \right) +%
\frac{3}{2}\Sigma ^{2}-\frac{2\Theta ^{2}}{3}=\frac{2{\mathcal{K}}}{a_{2}^{2}%
}\ ,  \label{2TimesGaussianCurv}
\end{equation}
where in the last equality the scale factor of the 2-sheets $a_2=a_2(t)$ has
been introduced and where ${\mathcal{K}}$ takes the values $\pm 1$ or 0
according to the geometry of the 2-sheets: sphere, pseudo-sphere, or flat.
Taking the time derivative of $\mathcal{R}$, and using Equations (\ref%
{continuity})--(\ref{Sigmadot}), one finds 
\begin{equation}
\mathcal{\dot{R}}=\left( \Sigma -\frac{2\Theta }{3}\right) \mathcal{R}\ ~,
\label{Kevo}
\end{equation}%
and hence one of the evolution Equations (\ref{continuity})--(\ref{Sigmadot}%
) can be replaced by Equation (\ref{Kevo}). According to the sign of ${%
\mathcal{R}}$, different types of solutions are obtained. For $\mathcal{R}>0$
one gets the Kantowski-Sachs cosmologies, which we studied in 
\cite{GWKS}. If $\mathcal{R}<0$ the spacetimes are of Bianchi type III. For $%
\mathcal{R}=0$ there are solutions of Bianchi type I/VII$_{0}$, including
the flat Friedmann universe. Since Equation~(\ref{2TimesGaussianCurv})
determines one of the quantities algebraically, one of the evolution
equations can be dropped. This~is due to the fact that the time derivative
of Equation (\ref{2TimesGaussianCurv}) will be identically satisfied due to
the evolution Equations (\ref{continuity})--(\ref{Sigmadot}).

The line-element can for the different values of ${\mathcal{K}}$ be written
as%
\begin{equation}  \label{metric}
ds^{2}=-dt^{2}+a_{1}^{2}\left( t\right) dz^{2}+a_{2}^{2}\left( t\right)
\left( d\vartheta ^{2}+f_{\mathcal{K}}(\vartheta) d\varphi ^{2}\right) \ ,
\end{equation}%
where $f_1(\vartheta)=\sin ^{2}\vartheta$, $f_{-1}(\vartheta)=\sinh
^{2}\vartheta$, and $f_0(\vartheta)=1$ (or alternatively $f_0=\vartheta$).
For ${\mathcal{K}}=1$ the 2-sheets are spheres and $\vartheta$ and $\varphi$
the usual spherical coordinates, but for ${\mathcal{K}}=-1$ and 0 the
2-sheets can be taken to be open and infinite with the topology of $R^2$.
The coordinates are dimensionless and hence the scale factors carry the
dimension of length (or time since $c=1$). The 4-velocity of comoving
observers is $u=\partial /\partial t$ and the direction of anisotropy is $%
n=a_{1}^{-1}\partial /\partial z$, which due to symmetry and normalisation
satisfies \cite{LRS}:%
\begin{equation}
\hat{n}_{b}\equiv n^{a}D_{a}n_{b}=0\ \ ,\ \ \ \dot{n}_{\bar{b}}=0\ \ .
\label{equation21}
\end{equation}

In terms of of the scale factors $a_1$ and $a_2$ in (\ref{metric}), the
expansion and scalar part of the shear take the~values 
\begin{equation}
\Theta =\frac{\dot{a}_{1}}{a_{1}}+2\frac{\dot{a}_{2}}{a_{2}}\ ,
\label{Thetabackgroundexp}
\end{equation}%
\begin{equation}
\Sigma =\frac{2}{3}\left( \frac{\dot{a}_{1}}{a_{1}}-\frac{\dot{a}_{2}}{a_{2}}%
\right) \; .  \label{Sigmabackgroundexp}
\end{equation}

\subsubsection{Homogeneous and Hypersurface Orthogonal LRS II Metrics with $%
\protect\phi\neq 0$}

There are also solutions with ${\mathcal{R}}=0$ and $\Sigma ={\mathcal{E}}=0$%
. For these the sheet expansion $\phi $ is in general nonzero and the system
is given by Equations (\ref{continuity}) and (\ref{expansiondot}) plus the
constraint 
\begin{equation}
\mu +\Lambda -\frac{1}{3}\Theta ^{2}+\frac{3}{4}\phi ^{2}=0\,.
\end{equation}

If $\phi \neq 0$, these are the negatively curved Friedmann models of
Bianchi type V, whereas $\phi =0$ gives the flat Friedmann models which are
covered by the sub-class in Section \ref{sectionphi0}. For the negatively
curved Friedmann models the metric can be given by 
\begin{equation}
ds^{2}=-dt^{2}+a^{2}\left[ d\tilde{z}^{2}+e^{-2\tilde{z}}\left(
dx^{2}+dy^{2}\right) \right] ~.\,  \label{metric20}
\end{equation}

\section{Vorticity-Free, Perfect Fluid Perturbations of Homogeneous and
Orthogonal LRS~II~Cosmologies}

\label{sectionperturb}

The analysis of perturbations on Kantowski-Sachs backgrounds in 
\cite{GWKS} will here be extended to all homogeneous and hypersurface
orthogonal LRS class II backgrounds except for the hyperbolic Friedmann
models. As for Kantowski-Sachs we will assume that the
perturbations are irrotational, i.e., that $\Omega=\Omega^a=0$, and also
that the perturbed spacetime is described by a perfect fluid. The frame is
partly fixed by choosing the preferred timelike vector $u^a$ to be the
4-velocity of the fluid also in the perturbed spacetime. Since the preferred
direction $n^a$ is not kept for the perturbed spacetime, we~choose to fix
its direction by choosing $a^a=0$, meaning that the acceleration only has a
component in the $n^a$-direction. For more details on the fixing of frame,
see \cite{GWKS}.

The choice of frame does not completely fix the mapping between the
perturbed and background spacetimes \cite{cov1}, but according to the Stewart-Walker 
lemma \cite{StewartWalker} variables which vanish on the
background are gauge-invariant. Hence, for the generic case with $\phi=0$ on
the background, we will replace the nonzero quantities on the background, $%
\mu$, $p$, $\Theta$, $\Sigma$ and ${\mathcal{E}}$ with their gradients 
\begin{equation}  \label{newvariables}
\mu_a=\delta_a\mu \, ,\;\; p_a=\delta_a p \, ,\;\; W_a=\delta_a\Theta \,
,\;\; V_a=\delta_a \Sigma \, ,\;\; X_a=\delta_a{\mathcal{E}} \, .
\end{equation}

As was shown in \cite{GWKS} the hat derivatives, $\widehat\Theta$ etc., are
determined in terms of the 2-gradients, $W_a$~etc., when the vorticity
vanishes (see also Section \ref{sectionhatharmonic}). The first-order
variables which vanish on the background are then 
\begin{equation}
S^{(1)}\equiv \left\{ X_{a},V_{a},W_{a},\mu _{a},p_{a},{\mathcal{A}},{%
\mathcal{A}}_{a},\Sigma _{a},\Sigma _{ab},{\mathcal{E}}_{a},{\mathcal{E}}%
_{ab},{\mathcal{H}},{\mathcal{H}}_{a},{\mathcal{H}}_{ab},a_{b},\alpha_a,\phi
,\xi ,\zeta _{ab}\right\} ~.
\end{equation}

These are now subject to the Ricci identities for $u^a$ and $n^a$ and
Bianchi identities, giving evolution equations along $u^a$, propagation
equations along $n^a$ and constraints. The exact, non-perturbative system of
equations for a 1 + 1 + 2 split of spacetime can be found in \cite{1+1+2}.
For first-order perturbations the set of equations in terms of the new
variables in Equation (\ref{newvariables}) was derived in \cite{GWKS}. This
was done for Kantowski-Sachs backgrounds, but this set is
actually valid for all hypersurface orthogonal homogeneous LRS class II
spacetimes with $\phi=0$, the differences lying in the zeroth order
coefficients for different backgrounds. For completeness we quote the result
from \cite{GWKS} in Appendix \ref{evolution0}.

For the special case ${\mathcal{R}}={\mathcal{E}}=\Sigma =0$ and $\phi \neq
0 $ we should in a similar way use the variable $\phi _{a}=\delta _{a}\phi $%
, whereas ${\mathcal{E}}$ and $\Sigma $ now are of first order. For the
modified system see Appendix \ref{evolution}.

\subsection{Harmonic Expansion}

\label{sectionharmonic}

For the metrics given by Equation (\ref{metric}), where $\phi =0$, the wave
equation on scalars 
\begin{equation}
\nabla ^{2}\Psi \equiv g^{ab}\nabla _{a}\nabla _{b}\Psi =0
\end{equation}%
is separable by applying the following harmonic expansion 
\begin{equation}
\Psi =\displaystyle\sum\limits_{k_{\parallel },k_{\perp }}\Psi
_{k_{\parallel }k_{\perp }}^{S}\ P^{k_{\parallel }}\ Q^{k_{\perp }}\ ,
\end{equation}%
where the coefficients $\Psi _{k_{\parallel }k_{\perp }}^{S}$ depend solely
on time (see, e.g., \cite{1+1+2,GWKS,KASAscalar}). The function $%
P^{k_{\parallel }}$ is the eigenfunction of the Laplacian $\widehat{\Delta }%
=n^{a}\nabla _{a}n^{b}\nabla _{b}$ and it is constant on the $z=const$
hypersurfaces 
\begin{equation}
\widehat{\Delta }P^{k_{\parallel }}=-\frac{k_{\parallel }^{2}}{a_{1}^{2}}%
P^{k_{\parallel }}\ ,\ \delta _{a}P^{k_{\parallel }}=\dot{P}^{k_{\parallel
}}=0\ .
\end{equation}

Here, $k_{\parallel }$ are the dimensionless constant comoving wave numbers
in the direction of anisotropy, and $a_{1}$ is the scale factor in this
direction. The physical wave numbers are given by $k_{\parallel }/a_{1}$.

Similarly, harmonics $Q^{k_\perp}$ are introduced on the 2-sheets as
eigenfunctions to the two-dimensional Laplace-Beltrami operator 
\cite{1+1+2}: 
\begin{equation}  \label{Beltrami}
\delta^2 Q^{k_\perp}=-\frac{k_\perp^2}{a_2^2} Q^{k_\perp}\, ,\ \widehat{Q}%
^{k_{\perp }}=\dot{Q}^{k_{\perp }}=0\ .
\end{equation}

Here, $\delta ^{2}=\delta _{a}\delta ^{a}$, $a_{2}$ is the scale factor of
the 2-sheets, and $k_\perp$ are the dimensionless comoving wavenumbers along
the 2-sheets.

When ${\mathcal{R}}>0$ the 2-sheets are spheres and the harmonics can be
represented by the usual spherical harmonics $Y_{l}^{m}$ 
\begin{equation}
\delta ^{2}Y_{l}^{m}=-\frac{l(l+1)}{a_{2}^{2}}Y_{l}^{m}~,~\widehat{Y}%
_{l}^{m}=\dot{Y}_{l}^{m}=0~,
\end{equation}%
with $k_{\perp }^{2}=l(l+1)$. Here $l=0,1,2,...$, and for a given $l$ value
the index $m$ runs from $-l$ to $l$. The~index $m$ does not occur in the
equations governing the perturbations due to the background
spacetime~symmetries.

For ${\mathcal{R}}\leq 0$, when the 2-sheets are open, the $k_\perp$ are not
discrete and may take any real values. For~${\mathcal{R}}=0$ the
eigenfunctions can be represented by plane waves.

Vectors and tensors can be also expanded in harmonics by introducing vector
and tensor harmonics \cite{Schperturb,Schperturb2,LRSIItensor}. The even
(electric) and odd (magnetic) parity vector harmonics are%
\begin{equation}
Q_{a}^{k_{\perp }}=a_{2}\delta _{a}Q^{k_{\perp }}\ ,\ \ \overline{Q}%
_{a}^{k_{\perp }}=a_{2}\varepsilon _{ab}\delta ^{b}Q^{k_{\perp }}\ ,
\end{equation}%
and the vector $\Psi _{a}$ can be expanded as%
\begin{equation}
\Psi _{a}=\displaystyle\sum\limits_{k_{\parallel },k_{\perp
}}P^{k_{\parallel }}\ \left( \Psi _{k_{\parallel }k_{\perp
}}^{V}Q_{a}^{k_{\perp }}+\overline{\Psi }_{k_{\parallel }k_{\perp }}^{V} 
\overline{Q}_{a}^{k_{\perp }}\right) \ .  \label{harmexpV0}
\end{equation}%
when the comoving wavenumbers take continous values, the sums are changed to
integrals with a~convenient normalization factor. For $Q_{a}^{k_{\perp }}$
and $\overline{Q}_{a}^{k_{\perp }}$ 
\begin{equation}
\varepsilon^{ab}\delta_a Q_{b}^{k_{\perp }}=0 \quad \hbox{and} \quad
\delta^a \overline{Q}_{a}^{k_{\perp }}=0
\end{equation}
hold, respectively. This corresponds to the fact that a generic vector can
be written as the sum of one curl-free and one divergence-free vector.

Similarly, the even and odd tensor harmonics are%
\begin{equation}
Q_{ab}^{k_{\perp }}=a_{2}^{2}\delta _{\{a}\delta _{b\}}Q^{k_{\perp }}\ ,\ 
\overline{Q}_{ab}^{k_{\perp }}=a_{2}^{2}\varepsilon _{c\{a}\delta ^{c}\delta
_{b\}}Q^{k_{\perp }}\ ,
\end{equation}%
and the tensor $\Psi _{ab}$ can be expanded as%
\begin{equation}
\Psi _{ab}=\displaystyle\sum\limits_{k_{\parallel },k_{\perp
}}P^{k_{\parallel }}\ \left( \Psi _{k_{\parallel }k_{\perp
}}^{T}Q_{ab}^{k_{\perp }}+\overline{\Psi }_{k_{\parallel }k_{\perp }}^{T} 
\overline{Q}_{ab}^{k_{\perp }}\right) \ .
\end{equation}

Note that for vectors ${\Psi }_a$ and tensors ${\Psi }_{ab}$ which are odd
by definition, the r\^oles of quantities without and with an overbar, e.g., $%
\overline{\Psi}_a$ , are interchanged. For example, for the magnetic part of
the Weyl tensor, where the three-dimensional volume element occurs in its
definition,  ${\mathcal{H}} _{k_{\parallel }k_{\perp}}^{T}$ belongs to the
odd sector, whereas $\overline{\mathcal{H}} _{k_{\parallel }k_{\perp}}^{T}$
belongs to the even sector.

Some useful relations involving the vector and tensor harmonics are listed
in Appendix \ref{harmonics}. For~different types of harmonics used in
relativity and cosmology see, for example, \cite%
{Challinor2,Gebbie1,Thorne,Harrison}.

\subsubsection{Harmonics When $\protect\phi\neq 0$}

In Bianchi V models with metric 
\begin{equation}
ds^{2}=-dt^{2}+\frac{a^{2}}{z^{2}}\left[ dz^{2}+\left( dx^{2}+dy^{2}\right) %
\right] ~,  \label{metric2}
\end{equation}%
where $z$ was introduced as $z=e^{\tilde{z}}$, the source free wave equation
for a scalar is: 
\begin{eqnarray}
0 &=&\nabla ^{2}\Psi =-\ddot{\Psi}-\Theta \Psi +D^{2}\Psi  \notag \\
&=&-\ddot{\Psi}-\Theta \Psi +\delta ^{2}\Psi +\phi \widehat{\Psi }+\widehat{%
\Delta }\Psi ~,
\end{eqnarray}%
with%
\begin{equation}
D^{2}\Psi =h^{ab}D_{a}D_{b}\Psi ~.
\end{equation}

In order to separate the time and spatial dependence of $\Psi $, we expand
it in harmonics obeying%
\begin{equation}
D^{2}Q^{k}=-\frac{k^{2}}{a^{2}}Q^{k}~,~\dot{Q}^{k}=0\ ,  \label{3dharm}
\end{equation}%
where $k$ is a real number. These harmonics can be built as follow. Equation
(\ref{Beltrami}) is modified to 
\begin{equation}
\delta ^{2}Q^{k_{\perp }}=-k_{\perp }^{2}\frac{z^{2}}{a^{2}}Q^{k_{\perp
}}\,,\ \widehat{Q}^{k_{\perp }}=\dot{Q}^{k_{\perp }}=0\ ,  \label{Beltrami2}
\end{equation}%
where $k_{\perp }$ is real and $Q^{k_{\perp }}$ can be represented by plane
waves:%
\begin{equation}
Q^{k_{\perp }}=e^{ik_{\perp }\left( x+y\right) }~.
\end{equation}

The expansion of $Q^{k}$ in harmonics $Q^{k_{\perp }}$ is given by%
\begin{equation}
Q^{k}\left( x,y,z\right) =\sum_{k_{\perp }}P^{k,k_{\perp }}\left( z\right)
Q^{k_{\perp }}\left( x,y\right) ~,
\end{equation}%
where the sum stands for a conveniently normalized integration with respect
to $k_{\perp }$ and $P^{k,k_{\perp }}\left( z\right) $~satisfies 
\begin{equation}
\left( z^{2}\frac{d^{2}}{dz^{2}}-z-k_{\perp }^{2}z^{2}\right) P^{k,k_{\perp
}}\left( z\right) =-k^{2}P^{k,k_{\perp }}\left( z\right) ~.  \label{Bess}
\end{equation}

This equation was derived from (\ref{3dharm}) and (\ref{Beltrami2}) and by
using $\phi =-2/a$ \cite{1+1+2}. The regular solution of~(\ref{Bess}) is%
\begin{equation}
P^{k,k_{\perp }}\left( z\right) =z^{3/2}K_{\nu }\left( k_{\perp }z\right) ~,
\end{equation}%
where $K_{\nu }$ is the modified Bessel functions of the second kind with%
\begin{equation}
\nu =\sqrt{1-k^{2}}~.
\end{equation}

A scalar occurring at the first order in the perturbed spacetime can be
expanded as%
\begin{equation}
\Psi =\sum_{k}\Psi _{k}^{S}\left( t\right) Q^{k}\left( x,y,z\right)
=\sum_{k,k_{\perp }}\Psi _{k}^{S}\left( t\right) P^{k,k_{\perp }}\left(
z\right) Q^{k_{\perp }}\left( x,y\right) ~.
\end{equation}

Nevertheless, this expansion shows that a 1 + 3 covariant approach is more
convenient in this case than the 1 + 1 + 2. The function $P^{k,k_{\perp
}}\left( z\right) $ depend on both separation constants $k$, $k_{\perp }$.
This is because $\delta ^{2}$-derivative carries a $z^{2}$ factor in
Equation (\ref{Beltrami2}).

For suitable three-dimensional harmonics see, for example, \cite%
{cov3,Challinor,Challinor2,perturb4,Gebbie1,EMM,Ullrich} and for 1 + 3
analysis of the Friedmann models see, e.g., \cite%
{cov1,DunsbyBassettEllis,Challinor,perturb4,Gebbie2,Marteens,Tsagas}.

\subsubsection{Relations between Harmonic Coefficients}

\label{sectionhatharmonic}

As was shown in \cite{GWKS}, on using the commutation relation (\ref%
{scalcomm4}) and the property (\ref{vectspherid4}) of the vector harmonics
and assuming vanishing vorticity, it follows that odd parts of the gradients
of the scalars $S^{(0)}=\{\Sigma ,\Theta ,\mu ,p,{\mathcal{E}}\}$ defined in
Equation (\ref{newvariables}) vanish: 
\begin{equation}
\overline{\mu }_{k_{\parallel }k_{\perp }}^{V}=\overline{X}_{k_{\parallel
}k_{\perp }}^{V}=\overline{V}_{k_{\parallel }k_{\perp }}^{V}=\overline{W}%
_{k_{\parallel }k_{\perp }}^{V}=\overline{p}_{k_{\parallel }k_{\perp
}}^{V}=0\ .  \label{OmegaSrel}
\end{equation}

It was also shown that the harmonic coefficients of the hat derivatives of
the objects in $S^{(0)}$ can be expressed in terms of the coefficients of
the vectors in Equation (\ref{newvariables}). Denoting an object in $S^{(0)}$
by $G$ its hat derivative can be expanded as 
\begin{equation}
\widehat{G}=\displaystyle\sum\limits_{k_{\parallel }k_{\perp }}\widetilde{G}%
_{k_{\parallel }k_{\perp }}^{S}\ P^{k_{\parallel }}\ Q^{k_{\perp }}\ ,
\label{harmexp2}
\end{equation}%
and its 2-gradients $G_{a}\equiv \delta _{a}G$ as 
\begin{equation}
G_{a}=\displaystyle\sum\limits_{k_{\parallel },k_{\perp }}P^{k_{\parallel
}}\ \left( G_{k_{\parallel }k_{\perp }}^{V}Q_{a}^{k_{\perp }}+\overline{G}%
_{k_{\parallel }k_{\perp }}^{V}\overline{Q}_{a}^{k_{\perp }}\right) \ .
\label{harmexpV}
\end{equation}

From (\ref{scalcomm3}) it then follows that 
\begin{equation}
\widetilde{G}_{k_{\parallel }k_{\perp }}^{S}=\frac{ik_{\parallel }a_{2}}{%
a_{1}}G_{k_{\parallel }k_{\perp }}^{V}~,\   \label{harmexpsrel}
\end{equation}%
if $\Omega _{a}$ and $\phi $ vanish to the zeroth order.

\subsection{Evolution Equations for the Case $\protect\phi = 0$}

\label{sectionevolution}

The evolution equations, propagation equations and constraints given in
Appendix \ref{evolution0} can be expanded in harmonics. This will result in
time evolution equations and constraints for the harmonic coefficients $%
\mathcal{A}_{k_{\parallel }k_{\perp }}^{S}$, $\mathcal{H}_{k_{\parallel
}k_{\perp }}^{S} $, $\!\phi _{k_{\parallel }k_{\perp }}^{S}$, $\xi
_{k_{\parallel }k_{\perp }}^{S}$, $\mu _{k_{\parallel }k_{\perp }}^{V}$, $%
p_{k_{\parallel }k_{\perp }}^{V}$, $\mathcal{A}_{k_{\parallel }k_{\perp
}}^{V}$, $\overline{\mathcal{A}}_{k_{\parallel }k_{\perp }}^{V}$, $%
V_{k_{\parallel }k_{\perp }}^{V}$, $W_{k_{\parallel }k_{\perp }}^{V}$, $%
X_{k_{\parallel }k_{\perp }}^{V}$, $\Sigma _{k_{\parallel }k_{\perp }}^{V}$, 
$\overline{\Sigma }_{k_{\parallel }k_{\perp }}^{V}$, $a_{k_{\parallel
}k_{\perp }}^{V}$, $\overline{a}_{k_{\parallel }k_{\perp }}^{V}$, $\alpha
_{k_{\parallel }k_{\perp }}^{V}$, $\overline{\alpha }_{k_{\parallel
}k_{\perp }}^{V}$, $\mathcal{E}_{k_{\parallel }k_{\perp }}^{V}$, $\overline{%
\mathcal{E}}_{k_{\parallel }k_{\perp }}^{V}$, $\mathcal{H}_{k_{\parallel
}k_{\perp }}^{V} $, $\overline{\mathcal{H}}_{k_{\parallel }k_{\perp }}^{V}$, 
$\Sigma _{k_{\parallel }k_{\perp }}^{T}$, $\overline{\Sigma }_{k_{\parallel
}k_{\perp }}^{T}$, $\zeta _{k_{\parallel }k_{\perp }}^{T}$, $\overline{\zeta 
}_{k_{\parallel }k_{\perp }}^{T}$, $\mathcal{E}_{k_{\parallel }k_{\perp
}}^{T}$, $\overline{\mathcal{E}}_{k_{\parallel }k_{\perp }}^{T}$, $\mathcal{H%
}_{k_{\parallel }k_{\perp }}^{T}$, and $\overline{\mathcal{H}}_{k_{\parallel
}k_{\perp }}^{T}$ . It follows that $\overline{\mathcal{A}}_{k_{\parallel
}k_{\perp }}^{V}=0$. The frame can then be fixed by requiring $a_a=0$, i.e., 
$a_{k_{\parallel}k_{\perp }}^{V}=\overline{a}_{k_{\parallel }k_{\perp
}}^{V}=0 $ which implies $\xi _{k_{\parallel}k_{\perp }}^{S}=0$. Finally, by
choosing a barytopic equation of state $p=p(\mu)$ we obtain $%
p_{k_{\parallel}k_{\perp }}^{V}=c_s^2\mu _{k_{\parallel }k_{\perp }}^{V}$ in
terms of the speed of sound squared $c_s^2=dp/d\mu$. Of the remaining 24
harmonic coefficients 18 can be solved for algebraically in terms of the six
coefficients $\mu _{k_{\parallel }k_{\perp }}^{V}$, $\Sigma _{k_{\parallel
}k_{\perp }}^{T}$, $\mathcal{E}_{k_{\parallel }k_{\perp}}^{T}$, $\overline{%
\mathcal{E}}_{k_{\parallel }k_{\perp }}^{T}$, $\mathcal{H}_{k_{\parallel
}k_{\perp }}^{T} $ and $\overline{\mathcal{H}}_{k_{\parallel }k_{\perp
}}^{T} $ (see Appendix \ref{harmoniccoefficients}). The remaining system for
the six harmonic coefficients decouple into two systems, one for the two
coefficients $\overline{\mathcal{E}}_{k_{\parallel }k_{\perp }}^{T}$ and $%
\mathcal{H}_{k_{\parallel }k_{\perp }}^{T}$ and one for the remaining four
coefficients $\mu _{k_{\parallel }k_{\perp }}^{V}$, $\Sigma _{k_{\parallel
}k_{\perp }}^{T}$, $\mathcal{E}_{k_{\parallel }k_{\perp}}^{T}$ and $%
\overline{\mathcal{H}}_{k_{\parallel }k_{\perp }}^{T}$

\subsubsection{System for $\overline{\mathcal{E}}_{k_{\parallel }k_{\perp
}}^{T}$ and $\mathcal{H}_{k_{\parallel }k_{\perp }}^{T}$}

It turns out that $\overline{\mathcal{E}}_{k_{\parallel }k_{\perp }}^{T}$
and $\mathcal{H}_{k_{\parallel }k_{\perp }}^{T}$ decouple from the other
coefficients. They satisfy the following system (note that both are of odd
parity): 
\begin{equation}
\dot{\overline{\mathcal{E}}}_{k_{\parallel }k_{\perp }}^{T}\!\!\!=-\frac{3}{2%
}\!\left( F\!+\!\Sigma D\!\right) \!\overline{\mathcal{E}}_{k_{\parallel
}k_{\perp }}^{T}\!\!\!+\frac{ik_{\parallel }}{a_{1}}\left( 1-D\right) 
\mathcal{H}_{k_{\parallel }k_{\perp }}^{T},  \label{40}
\end{equation}%
\begin{equation}
\dot{\mathcal{H}}_{k_{\parallel }k_{\perp }}^{T}=-\frac{a_{1}}{%
2ik_{\parallel }}\left( \frac{2k_{\parallel }^{2}}{a_{1}^{2}}-CB+9\Sigma
E\right) \overline{\mathcal{E}}_{k_{\parallel }k_{\perp }}^{T}-\frac{3}{2}%
\left( 2E+F\right) \mathcal{H}_{k_{\parallel }k_{\perp }}^{T}\ ,  \label{50}
\end{equation}%
where%
\begin{equation}
B\equiv \frac{2k_{\parallel }^{2}}{a_{1}^{2}}+\frac{k_{\perp }^{2}}{a_{2}^{2}%
}+\frac{9\Sigma ^{2}}{2}+3\mathcal{E}=\frac{2k_{\parallel }^{2}}{a_{1}^{2}}-%
\frac{{\mathcal{R}}a_2^2-k_{\perp }^{2}}{a_{2}^{2}}+3\Sigma \left( \Sigma +%
\frac{\Theta }{3}\right) \ ,  \label{Bdef}
\end{equation}%
\begin{equation}
CB\equiv\left( \frac{{\mathcal{R}}a_2^2-k_{\perp }^{2}}{a_{2}^{2}}+3\mathcal{%
E}\right)= \Sigma\left(\Theta-\frac{3}{2}\Sigma\right)-\frac{k_\perp^2}{a_2^2%
} \ ,  \label{Cdef}
\end{equation}
\begin{equation}
DB\equiv {\mathcal{R}}-\frac{k_\perp ^2}{a_2^2}+3{\mathcal{E}}+\mu+p=
\Sigma\left(\Theta-\frac{3}{2}\Sigma\right)-\frac{k_\perp ^2}{a_2^2}+\mu+p\ ,
\end{equation}%
\begin{equation}
EB\equiv \frac{\Sigma }{2}\left( CB-{\mathcal{E}}\right) +\frac{\Theta 
\mathcal{E}}{3}=\left(\frac{2}{3}\Theta^2-2(\mu+\Lambda)+\Sigma\left(\Theta-%
\frac{3}{2}\Sigma\right) \right)\left(\Theta-\frac{3}{2}\Sigma\right)-\Sigma%
\frac{k_\perp ^2}{2a_2^2} \ ,
\end{equation}%
\begin{equation}
F\equiv \Sigma +\frac{2\Theta }{3}\ .
\end{equation}

The system takes the same form as for the Kantowski-Sachs
background \cite{GWKS}, but note that the functions $B$, $C$ etc. are
slightly differently defined in terms of the curvature ${\mathcal{R}}$ of
the 2-sheets and also that the solutions from Equations (\ref{continuity})--(%
\ref{Sigmadot}), (\ref{Thetabackgroundexp}) and (\ref{Sigmabackgroundexp})
for the scale factors and kinematic quantities will be different for
different values of ${\mathcal{R}}$, given by Equation (\ref%
{2TimesGaussianCurv}).

The system can also be written as two decoupled second-order wave equations
with damping as 
\begin{equation}
\ddot{\overline{\mathcal{E}}}_{k_{\parallel }k_{\perp }}^{T}\!\!\!+q_{%
\overline{\mathcal{E}}1}\dot{\overline{\mathcal{E}}}_{k_{\parallel }k_{\perp
}}^{T}+q_{\overline{\mathcal{E}}0}\overline{\mathcal{E}}_{k_{\parallel
}k_{\perp }}^{T}=0~,  \label{waveE1}
\end{equation}%
\begin{equation}
\ddot{\mathcal{H}}_{k_{\parallel }k_{\perp }}^{T}+q_{\mathcal{H}1}\!\dot{%
\mathcal{H}}_{k_{\parallel }k_{\perp }}^{T}+q_{\mathcal{H}0}\mathcal{H}%
_{k_{\parallel }k_{\perp }}^{T}=0\ ,  \label{waveH1}
\end{equation}

where 
\begin{equation}
\begin{array}{rlll}
2q_{\overline{\mathcal{E}}0} & = & \frac{1-D}{a_{1}}\left[ \frac{%
2k_{\parallel }^{2}}{a_{1}}+a_{1}\left( 9\Sigma E-BC\right) \right] +3\frac{d%
}{dt}\left( F\!+\!\Sigma D\!\right) \! &  \\ 
&  & -3\left( F\!+\!\Sigma D\!\right) \left( \frac{d}{dt}\ln \frac{1-D}{a_{1}%
}-\frac{3}{2}\left( 2E+F\right) \right) ~, &  \\ 
&  &  & 
\end{array}%
\end{equation}
\begin{equation}
q_{\overline{\mathcal{E}}1}=\frac{3}{2}\!\left( 2E+2F\!+\!\Sigma D\!\right) -%
\frac{d}{dt}\ln \frac{1-D}{a_{1}}~,
\end{equation}%
\begin{equation}
\begin{array}{rlll}
2q_{\mathcal{H}0} & = & \frac{1-D}{a_{1}}\left[ \frac{2k_{\parallel }^{2}}{%
a_{1}}+a_{1}\left( 9\Sigma E-BC\right) \right] -3\left( 2E+F\right) \frac{d}{%
dt}\ln \left[ \frac{2k_{\parallel }^{2}}{a_{1}}+a_{1}\left( 9\Sigma
E-BC\right) \right] &  \\ 
&  & +\!\frac{9}{2}\left( 2E+F\right) \left( F\!+\!\Sigma D\!\right) +3\frac{%
d}{dt}\left( 2E+F\right) ~, &  \\ 
&  &  & 
\end{array}%
\end{equation}%
\begin{equation}
q_{\mathcal{H}1}=\frac{3}{2}\!\left( 2E+2F\!+\!\Sigma D\!\right) -\frac{d}{dt%
}\ln \left[ \frac{2k_{\parallel }^{2}}{a_{1}}+a_{1}\left( 9\Sigma
E-BC\right) \right] ~.
\end{equation}

In the high-frequency limit the speed of propagation for these waves will
approach the speed of light and hence they can be interpreted as free
gravitational waves (see Section \ref{sectionhighfrequency}).

\subsubsection{System for $\Sigma _{k_{\parallel }k_{\perp }}^{T}$, $%
\mathcal{E} _{k_{\parallel }k_{\perp }}^{T}$, $\overline{\mathcal{H}}%
_{k_{\parallel }k_{\perp }}^{T}$ and $\protect\mu _{k_{\parallel }k_{\perp
}}^{V}$}

\label{sources}

The coefficients $\Sigma _{k_{\parallel }k_{\perp }}^{T}$, $\mathcal{E}%
_{k_{\parallel }k_{\perp }}^{T}$, $\overline{\mathcal{H}}_{k_{\parallel
}k_{\perp }}^{T}$ and $\mu _{k_{\parallel }k_{\perp }}^{V}$ form the
following system: 
\begin{equation}
\begin{array}{rlll}
\dot{\mu}_{k_{\parallel }k_{\perp }}^{V}\!\! & = & \left[ \frac{\Sigma }{2}%
\left( 1-3\frac{\mu +p}{B}\right) \!-\frac{4\Theta }{3}\right] \mu
_{k_{\parallel }k_{\perp }}^{V}\!+\frac{a_{2}}{2}\left( \mu +p\right) &  \\ 
&  & \times \left[ \left( 1-C\right) \!\left( B\Sigma _{k_{\parallel
}k_{\perp }}^{T}\!-3\Sigma \mathcal{E}_{k_{\parallel }k_{\perp }}^{T}\right)
+\frac{ik_{\parallel }}{a_{1}}(2-J)\overline{\mathcal{H}}_{k_{\parallel
}k_{\perp }}^{T}\right] ~, \label{eq11} &  \\ 
&  &  & 
\end{array}%
\end{equation}%
\begin{equation}
\dot{\Sigma}_{k_{\parallel }k_{\perp }}^{T}=-\frac{c_{s}^{2}}{a_{2}\left(
\mu +p\right) }\mu _{k_{\parallel }k_{\perp }}^{V}+\!\left( \Sigma -\frac{%
2\Theta }{3}\right) \Sigma _{k_{\parallel }k_{\perp }}^{T}-\mathcal{E}%
_{k_{\parallel }k_{\perp }}^{T}\ ,  \label{eq12}
\end{equation}%
\begin{equation}
\dot{\mathcal{E}}_{k_{\parallel }k_{\perp }}^{T}\!=\!\frac{3\Sigma }{2a_{2}B}%
\mu _{k_{\parallel }k_{\perp }}^{V}-\frac{\mu +p}{2}\Sigma _{k_{\parallel
}k_{\perp }}^{T}-\frac{3}{2}\left( F+\Sigma C\right) \mathcal{E}%
_{k_{\parallel }k_{\perp }}^{T}-\frac{ik_{\parallel }}{2a_{1}}(2-J)\overline{%
\mathcal{H}}_{k_{\parallel }k_{\perp }}^{T}\ ,  \label{eq13}
\end{equation}%
\begin{equation}
\dot{\overline{\mathcal{H}}}_{k_{\parallel }k_{\perp }}^{T}\!\!\!\!=-\frac{%
ik_{\parallel }}{a_{1}a_{2}B}\mu _{k_{\parallel }k_{\perp }}^{V}-\frac{3}{2}%
\left( \frac{M}{B}+F\right) \overline{\mathcal{H}}_{k_{\parallel }k_{\perp
}^{T}}-\frac{ik_{\parallel }}{a_{1}}\left( 1-C\right) \mathcal{E}%
_{k_{\parallel }k_{\perp }}^{T}\ .  \label{eq14}
\end{equation}

Here we have introduced the additional notations 
\begin{equation}
\begin{array}{rlll}
JB & \equiv & \frac{({\mathcal{R}}a_{2}^{2}-k_{\perp }^{2})k_{\perp
}^{2}a_{1}^{2}}{k_{\parallel }^{2}a_{2}^{4}}+2CB=\left( {\mathcal{R}}-\frac{%
k_{\perp }^{2}}{a_{2}^{2}}\right) \left( \frac{2k_{\parallel }^{2}}{a_{1}^{2}%
}+\frac{k_{\perp }^{2}}{a_{2}^{2}}\right) \frac{a_{1}^{2}}{k_{\parallel }^{2}%
}+6{\mathcal{E}} &  \\ 
& = & \frac{k_{\perp }^{2}a_{1}^{2}}{k_{\parallel }^{2}a_{2}^{2}}\left( {%
\mathcal{R}}-\frac{2k_{\parallel }^{2}}{a_{1}^{2}}-\frac{k_{\perp }^{2}}{%
a_{2}^{2}}\right) +2\Sigma \left( \Theta -\frac{3}{2}\Sigma \right) \,, & 
\\ 
&  &  & 
\end{array}%
\end{equation}%
\begin{equation}
M\equiv 2{\mathcal{E}}\left( \Sigma +\frac{\Theta }{3}\right) +\Sigma \frac{{%
\mathcal{R}}a_{2}^{2}-k_{\perp }^{2}}{a_{2}^{2}}~,\ 
\end{equation}%
where ${\mathcal{E}}$ is given by Equation (\ref{EWeyl}).

As for the Kantowski-Sachs case, from these one can derive
second-order wave-like equations for $\Sigma_{k_{\parallel }k_{\perp }}^{T}$%
, $\mathcal{E}_{k_{\parallel }k_{\perp }}^{T}$, and $\overline{\mathcal{H}}%
_{k_{\parallel }k_{\perp }}^{T}$ where the density gradient $%
\mu_{k_{\parallel }k_{\perp }}^{V}$ and its derivative act as source terms.
It is only in the high frequency limit that the second-order equations for $%
\mathcal{E}_{k_{\parallel }k_{\perp }}^{T}$ and $\overline{\mathcal{H}}%
_{k_{\parallel }k_{\perp }}^{T}$ decouple from the source terms, given by
the density gradient, and hence describe freely moving gravitational waves.

\subsection{High-Frequency Approximation}

\label{sectionhighfrequency}

In \cite{GWKS}, where the backgrounds were given by Kantowski-Sachs 
models, we studied the high-frequency limit (optical limit; see \cite%
{Isaacson1,Isaacson2}), of the propagation equations. For our quantities
this implies 
\begin{equation}
\frac{k_{\parallel }^{2}}{a_{1}^{2}},\;\frac{k_{\perp }^{2}}{a_{2}^{2}}\gg
\Theta ^{2},\;\Sigma ^{2},\;{\mathcal{E}},\;\mu ,\;p \, .
\end{equation}

Since in this limit the curvature of the 2-sheets becomes negligible the
resulting equations are identical in form to those for the Kantowski-Sachs 
backgrounds for all signs of ${\mathcal{R}}$, but the
zeroth-order factors $a_{1}$, $a_{2}$, $\Theta $ and $\Sigma $ of course are
different for different backgrounds.

For the uncoupled system of $\overline{\mathcal{E}}_{k_{\parallel }k_{\perp
}}^{T}$ and $\mathcal{H}_{k_{\parallel }k_{\perp }}^{T}$ the following
second-order wave equations with~damping 
\begin{equation}
\ddot{\overline{\mathcal{E}}}_{k_{\parallel }k_{\perp }}^{T}\!\!\!+q_{%
\overline{\mathcal{E}}1}\dot{\overline{\mathcal{E}}}_{k_{\parallel }k_{\perp
}}^{T}+\left( \frac{k_{\parallel }^{2}}{a_{1}^{2}}+\frac{k_{\perp }^{2}}{%
a_{2}^{2}}\right) \overline{\mathcal{E}}_{k_{\parallel }k_{\perp }}^{T}=0~,
\end{equation}%
\begin{equation}
\ddot{\mathcal{H}}_{k_{\parallel }k_{\perp }}^{T}+q_{\mathcal{H}1}\!\dot{%
\mathcal{H}}_{k_{\parallel }k_{\perp }}^{T}+\left( \frac{k_{\parallel }^{2}}{%
a_{1}^{2}}+\frac{k_{\perp }^{2}}{a_{2}^{2}}\right) \mathcal{H}_{k_{\parallel
}k_{\perp }}^{T}=0\ ,
\end{equation}%
where  
\begin{equation}
q_{\overline{\mathcal{E}}_{1}}=\frac{7\Theta }{3}+4\Sigma -3\Sigma \frac{%
k_{\perp }^{2}}{\frac{a_{2}^{2}}{a_{1}^{2}}k_{\parallel }^{2}+k_{\perp }^{2}}%
~,
\end{equation}%
\begin{equation}
q_{{\mathcal{H}}_{1}}=\frac{7\Theta }{3}+4\Sigma -6\Sigma \frac{k_{\perp
}^{2}}{\frac{2a_{2}^{2}}{a_{1}^{2}}k_{\parallel }^{2}+k_{\perp }^{2}}~,
\end{equation}%
are obtained. These are in the form 
\begin{equation}
\ddot{X}+2\zeta \Omega \dot{X}+\Omega ^{2}X=0\,,
\end{equation}%
where $\Omega $ is the undamped angular frequency and the actual angular
frequency is given by $\Omega \sqrt{1-\zeta ^{2}}$. The propagation speed of
the wave is 
\begin{equation}
c_{w}=\frac{\Omega }{k_{phys}}\sqrt{1-\zeta ^{2}}\quad \hbox{with}\quad
k_{phys}^{2}=\frac{k_{\parallel }^{2}}{a_{1}^{2}}+\frac{k_{\perp }^{2}}{%
a_{2}^{2}}\,.
\end{equation}

The propagation velocity hence goes as $1-\zeta ^{2}/2$ for relatively small
damping coefficients $\zeta$, and approaches the speed of light for large
frequencies. For the static case, when $\Theta =\Sigma =0$, the damping
would vanish and then the propagation velocity would be exactly the speed of
light. When the propagation is along the preferred direction, $k_{\perp }=0,$
and $q_{\overline{\mathcal{E}}_{1}}=q_{{\mathcal{H}}_{1}}$, resulting in the
same damping $\zeta _{||}$ for both variables $\overline{\mathcal{E}}%
_{k_{\parallel }k_{\perp }}^{T}$ and $\mathcal{H}_{k_{\parallel }k_{\perp
}}^{T}$. Hence, they have the same propagation velocity which differs from
the speed of light at the second-order in $\zeta _{||}$. However, when the
propagation is perpendicular to the preferred direction, $k_{\parallel
}^{2}=0$, then $q_{\overline{\mathcal{E}}_{1}}\neq q_{{\mathcal{H}}_{1}}$,
giving different dampings $\zeta _{\perp \overline{\mathcal{E}}}$ and $\zeta
_{\perp \mathcal{H}}$ for $\overline{\mathcal{E}}_{k_{\parallel }k_{\perp
}}^{T}$ and $\mathcal{H}_{k_{\parallel }k_{\perp }}^{T}$, respectively.
Therefore, the propagation velocities also differ at the second order in
damping coefficients. In addition, since $\zeta _{||}\neq \zeta _{\perp 
\overline{\mathcal{E}}}\neq \zeta _{\perp \mathcal{H}}$, the propagation
velocities are direction-dependent.

As mentioned in Section \ref{sources}, the second-order equations for ${%
\mathcal{E}}_{k_{\parallel }k_{\perp }}^{T}$ and $\overline{\mathcal{H}}%
_{k_{\parallel }k_{\perp }}^{T}$ also decouple from the density gradient in
the high frequency limit and are given by 
\begin{equation}
\ddot{\mathcal{E}}_{k_{\parallel }k_{\perp }}^{T}+q_{\mathcal{E}1}\dot{%
\mathcal{E}}_{k_{\parallel }k_{\perp }}^{T}+\left( \frac{k_{\parallel }^{2}}{%
a_{1}^{2}}+\frac{k_{\perp }^{2}}{a_{2}^{2}}\right) \mathcal{E}_{k_{\parallel
}k_{\perp }}^{T}=0~,
\end{equation}%
\begin{equation}
\ddot{\overline{\mathcal{H}}}_{k_{\parallel }k_{\perp }}^{T}+q_{\overline{%
\mathcal{H}}1}\dot{\overline{\mathcal{H}}}_{k_{\parallel }k_{\perp
}}^{T}+\left( \frac{k_{\parallel }^{2}}{a_{1}^{2}}+\frac{k_{\perp }^{2}}{%
a_{2}^{2}}\right) \mathcal{H}_{k_{\parallel }k_{\perp }}^{T}=0~,
\end{equation}%
with $q_{\mathcal{E}1}=q_{{\mathcal{H}}_{1}}$ and $q_{\overline{\mathcal{H}}%
1}=q_{\overline{\mathcal{E}}_{1}}$. Beause of the latter, the propagation
velocities of $\mathcal{E}_{k_{\parallel }k_{\perp }}^{T}$ and $\mathcal{H}%
_{k_{\parallel }k_{\perp }}^{T}$ and of $\overline{\mathcal{H}}%
_{k_{\parallel }k_{\perp }}^{T}$ and $\overline{\mathcal{E}}_{k_{\parallel
}k_{\perp }}^{T}$ coincide in the high-frequency limit and as before the
propagation velocity differs from the speed of light to the second order in
the damping parameter.

\subsection{Perturbations of the flat Friedmann models}

\label{sectionFriedmann}

Here we consider the flat Friedmann models as a check of the isotropic limit
of the general LRS~II case. In this case there is no preferred spatial
direction on the background, but as before, we fix the 1-direction by
choosing the acceleration to only have a 1-component, i.e., $a_a=0$, in the
perturbed spacetime. Tensor perturbations of the Friedmann cases, using the
1 + 3 covariant split, were studied in \cite{DunsbyBassettEllis} and we make
a comparison with their results.

The flat Friedmann models are given by ${\mathcal{E}} =\Sigma ={\mathcal{R}}%
=\phi =0$. Without loss of generality we can use $a_{1}=a_{2}\equiv a$. We
also introduce the notations 
\begin{equation}
k^{2}\equiv k_{\parallel }^{2}+k_{\perp }^{2}~,
\end{equation}%
and 
\begin{equation}
B_{0}=\frac{2k_{\parallel }^{2}}{a^{2}}+\frac{k_{\perp }^{2}}{a^{2}}~.
\end{equation}

The first system for $\overline{{\mathcal{E}}}_{k_{\parallel }k_{\perp
}}^{T} $ and $\mathcal{H}_{k_{\parallel }k_{\perp }}^{T}$ then reduces to 
\begin{equation}
\dot{\overline{\mathcal{E}}}_{k_{\parallel }k_{\perp }}^{T}\!\!\!=-\Theta \,%
\overline{\mathcal{E}}_{k_{\parallel }k_{\perp }}^{T}\!\!\!+\frac{%
ik_{\parallel }}{aB_{0}}\left( \frac{2k^{2}}{a^{2}}-\mu -p\right) \mathcal{H}%
_{k_{\parallel }k_{\perp }}^{T},  \label{4}
\end{equation}%
\begin{equation}
\dot{\mathcal{H}}_{k_{\parallel }k_{\perp }}^{T}=-\frac{a}{2ik_{\parallel }}%
B_{0}\overline{\mathcal{E}}_{k_{\parallel }k_{\perp }}^{T}-\Theta \mathcal{H}%
_{k_{\parallel }k_{\perp }}^{T}\ .  \label{5}
\end{equation}

As before, this can be written as second-order damped wave equations 
\begin{equation}  \label{EbarF}
\ddot{\overline{\mathcal{E}}}_{k_{\parallel }k_{\perp }}^{T}\!\!\!+ \left( 
\frac{7\Theta}{3}-N \right) \dot{\overline{\mathcal{E}}}_{k_{\parallel
}k_{\perp }}^{T}+ \left(\frac{k^2}{a^2}+\frac{2 }{3}\Theta^2+2(\Lambda-p)-N%
\Theta\right) \overline{\mathcal{E}}_{k_{\parallel }k_{\perp }}^{T}=0 \,
\end{equation}
and 
\begin{equation}  \label{HF}
\ddot{\mathcal{H}}_{k_{\parallel }k_{\perp }}^{T}+\frac{7}{3}\Theta\, \dot{%
\mathcal{H}}_{k_{\parallel }k_{\perp }}^{T}+\left( \frac{k^2}{a^{2}}+\frac{2 
}{3}\Theta^2+2(\Lambda-p)\right) \mathcal{H}_{k_{\parallel }k_{\perp
}}^{T}=0 \, .
\end{equation}

The term $N$ is given by 
\begin{equation}
N\equiv \frac{(\mu+p)\left(1+3c_s^2\right)}{3\left(\frac{2k^2}{a^2}%
-(\mu+p)\right)}\Theta \,
\end{equation}
and vanishes in the high frequency limit.

The second system becomes 
\begin{equation}  \label{eq11F}
\dot{\mu}_{k_{\parallel }k_{\perp }}^{V}=-\frac{4\Theta }{3}\mu
_{k_{\parallel }k_{\perp }}^{V}\!+\frac{k^{2}}{a}(\mu
+p)\Sigma_{k_{\parallel }k_{\perp }}^{T}-\frac{a^{2}(\mu +p)B_{0}}{%
2ik_{\parallel }}\overline{\mathcal{H}}_{k_{\parallel }k_{\perp }}^{T}~,
\end{equation}%
\begin{equation}
\dot{\Sigma}_{k_{\parallel }k_{\perp }}^{T}=-\frac{c_{s}^{2}}{a\left( \mu
+p\right) }\mu _{k_{\parallel }k_{\perp }}^{V}-\frac{2\Theta }{3}\Sigma
_{k_{\parallel }k_{\perp }}^{T}-\mathcal{E}_{k_{\parallel }k_{\perp }}^{T}\ ,
\label{eq12F}
\end{equation}%
\begin{equation}
\dot{\mathcal{E}}_{k_{\parallel }k_{\perp }}^{T}\!=-\frac{\mu +p}{2}\Sigma
_{k_{\parallel }k_{\perp }}^{T}-\Theta \mathcal{E}_{k_{\parallel }k_{\perp
}}^{T}+\frac{aB_{0}}{2ik_{\parallel }}\overline{\mathcal{H}}_{k_{\parallel
}k_{\perp }}^{T}\ ,  \label{eq13F}
\end{equation}%
\begin{equation}
\dot{\overline{\mathcal{H}}}_{k_{\parallel }k_{\perp }}^{T}\!\!\!\!=-\frac{%
ik_{\parallel }}{a^{2}B_{0}}\mu _{k_{\parallel }k_{\perp }}^{V}-\Theta 
\overline{\mathcal{H}}_{k_{\parallel }k_{\perp }}^{T}-\frac{2ik_{\parallel
}k^{2}}{a^{3}B_{0}}\mathcal{E}_{k_{\parallel }k_{\perp }}^{T}\ .
\label{eq14F}
\end{equation}

From these we obtain the following second-order wave equations for $%
\Sigma_{k_{\parallel }k_{\perp }}^{T}$ and ${\mathcal{E}}_{k_{\parallel
}k_{\perp }}^{T}$, again with the density fluctuations acting as source
terms, 
\begin{equation}
\ddot{\Sigma}_{k_{\parallel }k_{\perp }}^{T}+\frac{5\Theta}{3}\dot{\Sigma}%
_{k_{\parallel }k_{\perp }}^{T}+\left(\frac{k^2}{a^2}+\frac{\Theta^2}{6}+%
\frac{3}{2}(\Lambda-p)\right)\Sigma _{k_{\parallel }k_{\perp }}^{T}=\frac{%
\left( 1-c_{s}^{2}\right) }{a\left( \mu +p\right) }\dot{\mu}_{k_{\parallel
}k_{\perp }}^{V}+s_{\Sigma 0}\mu _{k_{\parallel }k_{\perp }}^{V}~,
\label{SigmaEqF}
\end{equation}%
\begin{equation}
\ddot{\mathcal{E}}_{k_{\parallel }k_{\perp }}^{T}\!+\left(\frac{7\Theta}{3}%
-N \right)\dot{\mathcal{E}}_{k_{\parallel }k_{\perp }}^{T}+ \left(\frac{k^2}{%
a^2}+\frac{2 }{3}\Theta^2+2(\Lambda-p)-N\Theta\right) \mathcal{E}%
_{k_{\parallel }k_{\perp }}^{T}=\frac{N}{a(\mu+p)}\dot{\mu}_{k_{\parallel
}k_{\perp }}^{V}+s_{\mathcal{E}0}\mu _{k_{\parallel }k_{\perp }}^{V}\ ,
\label{EEqF}
\end{equation}%
where 
\begin{equation}
s_{\Sigma 0}=\frac{\left(\left(\frac{4}{3}-\frac{5}{3} c_s^2
-c_s^4\right)\Theta- 2 c_s \dot c_s\right)}{a\left( \mu +p\right) } \quad %
\hbox{and} \quad s_{{{\mathcal{E}}0}}=\frac{c_s^2-1}{2a}+\frac{4\Theta N}{%
3a(\mu+p)}\, .
\end{equation}

The density gradient obeys the following second-order equation 
\begin{equation}  \label{densitywave}
\ddot{\mu}_{k_{\parallel }k_{\perp }}^{V}+\left(\frac{10}{3}%
+c_s^2\right)\theta \dot{\mu}_{k_{\parallel }k_{\perp }}^{V} +\left(\frac{%
c_s^2 k^2}{a^2}+\frac{11}{6}\theta^2+\frac{4 c_s^2}{3}\theta^2-\frac{5}{2}%
\left(p-\Lambda\right)\right) {\mu}_{k_{\parallel }k_{\perp }}^{V}=0
\end{equation}
and we see that the density perturbations propagate with the speed of sound
in the high frequency limit. Unlike the case with an anisotropic background,
we now obtain a second-order decoupled wave equation for $\overline{\mathcal{%
H}}_{k_{\parallel }k_{\perp }}^{T}$ 
\begin{equation}
\ddot{\overline{\mathcal{H}}}_{k_{\parallel }k_{\perp }}^{T}+\frac{7}{3}%
\Theta \,\dot{\overline{\mathcal{H}}}_{k_{\parallel }k_{\perp }}^{T}+\left( 
\frac{k^{2}}{a^{2}}+\frac{2}{3}\Theta ^{2}+2(\Lambda -p)\right) \overline{%
\mathcal{H}}_{k_{\parallel }k_{\perp }}^{T}=0\, .  \label{HbarF}
\end{equation}

Note that Equations (\ref{HF}) and (\ref{HbarF}) for ${\mathcal{H}}%
_{k_{\parallel }k_{\perp }}^{T}$ and $\overline{\mathcal{H}}_{k_{\parallel
}k_{\perp }}^{T}$, respectively, are identical. Similarly, the left-hand
sides of Equations (\ref{EbarF}) for $\overline{\mathcal{E}} _{k_{\parallel
}k_{\perp }}^{T}$ and (\ref{EEqF}) for $\mathcal{E} _{k_{\parallel }k_{\perp
}}^{T}$, respectively, are identical. The damping coefficients in the
second-order equations depend only on $k$, therefore the propagation
velocities of the perturbations are not direction-dependent. In addition the
propagation velocities at high frequencies approach the speed of light.
Because of the symmetries of Friedmann spacetimes, we can safely assume that 
$\mu _{k_{\parallel }k_{\perp }}^{V}$ depend only on $k$, and not separately
on $k_{\parallel }$ and $k_{\perp }$, which is also clear from Equation (\ref%
{densitywave}). Then the second-order equations for $\Sigma _{k_{\parallel
}k_{\perp }}^{T}$ and $\mathcal{E}_{k_{\parallel }k_{\perp }}^{T}$ also
depend only on $k$ like those governing $\overline{\mathcal{E}}%
_{k_{\parallel }k_{\perp }}^{T}$, ${\mathcal{H}}_{k_{\parallel }k_{\perp
}}^{T}$ and $\overline{\mathcal{H}}_{k_{\parallel }k_{\perp }}^{T}$.

In \cite{DunsbyBassettEllis} the 1 + 3 covariant split was used to study
tensorial perturbations of the Friedmann models. The pure tensor
perturbations are characterised by vanishing energy density gradients and
vorticity to the first order. Hence, when comparing our result with their we
must use $\mu_{k_{\parallel}k_{\perp}}^V=\dot
\mu_{k_{\parallel}k_{\perp}}^V=0$. Our~Equations (\ref{HF}) and (\ref{HbarF}%
) for the even and odd parts of the magnetic part of the Weyl tensor, $%
\overline{\mathcal{H}}_{k_{\parallel }k_{\perp }}^{T}$ and $\mathcal{H}%
_{k_{\parallel}k_{\perp }}^{T}$ , respectively, are then the same as their
Equation (20) for the flat Friedmann case. The~equation for $\Sigma
_{k_{\parallel }k_{\perp}}^{T}$ (\ref{SigmaEqF}) similarly corresponds to
their Equation (22){\ \footnote{{There seems to be a misprint in Equation}
(22) in \cite{DunsbyBassettEllis}. Probably the factor $(9\gamma -1)$ should
read $(10-9\gamma)$.}}.

The main difference between \cite{DunsbyBassettEllis} and our result is that
we obtain a second-order wave equation from Equation (\ref{EEqF}) for $%
\mathcal{E} _{k_{\parallel }k_{\perp }}^{T}$, whereas they needed a
third-order equation to decouple~$E_{ab}$. In~Equation~(15) in \cite%
{DunsbyBassettEllis} they give a second-order equation for the electric part
of the Weyl tensor, $E_{ab}$, with the shear $\sigma_{ab}$ acting as a
source term. From our Equation (\ref{eq11F}) we see that imposing $%
\mu_{k_{\parallel}k_{\perp}}^V=\dot \mu_{k_{\parallel}k_{\perp}}^V=0$ gives
a constraint between $\Sigma _{k_{\parallel }k_{\perp}}^{T}$ and $\overline{%
\mathcal{H}}_{k_{\parallel }k_{\perp }}^{T}$ 
\begin{equation}  \label{SigmaH}
\Sigma _{k_{\parallel }k_{\perp}}^{T}=\frac{a^3 B_0}{2k^2 ik_\parallel}%
\overline{\mathcal{H}}_{k_{\parallel }k_{\perp }}^{T} \, .
\end{equation}

This constraint is satisfied as is seen by differentiating it and
substituting Equations (\ref{eq12F})--(\ref{eq14F}). Hence,~the number of
independent variables in the system (\ref{eq11F})--(\ref{eq14F}) is reduced
from four to two by imposing that the density perturbations vanish. With the
help of Equation (\ref{eq13F}) now $\Sigma _{k_{\parallel }k_{\perp}}^{T}$
can be completely expressed in terms of $\mathcal{E} _{k_{\parallel
}k_{\perp }}^{T}$ and $\dot{\mathcal{E}} _{k_{\parallel }k_{\perp }}^{T}$ 
\begin{equation}  \label{SigmaE}
\Sigma _{k_{\parallel }k_{\perp}}^{T}=\frac{2a^2}{2k^2-(\mu+p)a^2}\left(\dot{%
\mathcal{E}} _{k_{\parallel }k_{\perp }}^{T} +\Theta {\mathcal{E}}
_{k_{\parallel }k_{\perp }}^{T}\right) \, .
\end{equation}

By using this result in their Equation (\ref{expansiondot}) in for the flat
Friedmann case, where simple plane waves can be used for the harmonics, we
obtain our Equation (\ref{EEqF}). The three-dimensional 
Equations (\ref{continuity0}) and (\ref{equation21}) 
in \cite{DunsbyBassettEllis} correspond to our Equations (\ref{SigmaH}) and (%
\ref{SigmaE}). Using the three-dimensional harmonics in \cite{Challinor2}
similar second-order equations for all values of ${\mathcal{K}}$ are
obtained (see Appendix \ref{App3dim}).


\section{Conclusions}

\label{conclusion}

A previous analysis of vorticity-free perturbations on Kantowski-Sachs 
backgrounds has been extended to all homogeneous and hypersurface
orthogonal LRS perfect fluids with the vanishing magnetic part of the Weyl
tensor except the hyperbolic and closed Friedmann models, which have been
studied elsewere \cite%
{cov1,DunsbyBassettEllis,Challinor,perturb4,Gebbie2,Marteens,Tsagas} using
the covariant 1 + 3 split approach. We find the same structure of the
evolution equations for the perturbations as in the case of Kantowski-Sachs. 
All harmonic coefficients can be determined in terms of a
subset containing only six coefficients. The evolution equations for these
decouple into one system for $\overline{\mathcal{E}}_{k_{\parallel }k_{\perp
}}^{T}$ and $\mathcal{H}_{k_{\parallel}k_{\perp }}^{T}$, representing
source-free gravitational degrees of freedom, and another for $\Sigma
_{k_{\parallel }k_{\perp }}^{T}$, $\mathcal{E} _{k_{\parallel }k_{\perp
}}^{T}$, $\overline{\mathcal{H}}_{k_{\parallel }k_{\perp }}^{T}$, and $\mu
_{k_{\parallel }k_{\perp }}^{V}$, which describes perturbations sourced by
the density gradient. Only in the high frequency limit do the second-order
wave equations for $\mathcal{E}_{k_{\parallel }k_{\perp }}^{T}$ and $%
\overline{\mathcal{H}}_{k_{\parallel }k_{\perp }}^{T}$ decouple from the
source terms. The analysis of propagation velocities in the high frequency
limit led to direction-dependent dispersion relations on
anisotropic~backgrounds.

We also studied perturbations on the flat Friedmann universe, which is the
isotropic limit of the considered class of backgrounds. Here the
second-order wave equation for $\overline{\mathcal{H}}_{k_{\parallel}k_{%
\perp }}^{T}$ decouples from the other coefficients, whereas $\mathcal{E}%
_{k_{\parallel }k_{\perp }}^{T}$ still is sourced by the density gradient.
The result is compared with an earlier study,  \cite{DunsbyBassettEllis},
where a 1 + 3 formalism was used to study pure tensor perturbations on
Friedmann backgrounds. The 1 + 1 + 2 covariant approach, together with a
harmonic decomposition, is also effective for this setup, because all
tensorial perturbations are  easily obtained as second-order differential
equations in contrast to the third-order equation more naturally appearing
for the electric part of Weyl tensor in the 1 + 3 covariant description.

\vspace{6pt}

\section*{Acknowledgements}

The work of ZK was partially supported by the UNKP-17-4 New National
Excellence Program of the Ministry of Human Capacities and partially  by the
Hungarian National Research, Development and Innovation Office (NKFI) in the
form of the grant 123996.

\appendix

\section{Commutation Relations}

\label{commutation} The commutation relations of covariant derivatives of
the scalar field $\Psi $ on hypersurface orthogonal and homogeneous LRS II
backgrounds are to the first order:%
\begin{equation}
\widehat{\dot{\Psi}}-\dot{\widehat{\Psi }}=-\mathcal{A}\dot{\Psi}+\left(
\Sigma +\frac{\Theta }{3}\right) \widehat{\Psi }\ ,
\end{equation}%
\begin{equation}
\delta _{a}\dot{\Psi}-N_{a}^{\,\,\,b}\left( \delta _{b}\Psi \right) ^{\cdot
}=-\mathcal{A}_{a}\dot{\Psi}-\!\!\frac{1}{2}\left( \Sigma -\frac{2\Theta }{3}%
\right) \delta _{a}\Psi \ ,
\end{equation}%
\begin{equation}
\delta _{a}\widehat{\Psi }-N_{a}^{\,\,\,b}\left( \widehat{\delta _{b}\Psi }%
\right) =-2\varepsilon _{ab}\Omega ^{b}\dot{\Psi}+\frac{1}{2}%
\phi\delta_a\Psi\ ,  \label{scalcomm3}
\end{equation}%
\begin{equation}
\delta _{\lbrack a}\delta _{b]}\Psi =\varepsilon _{ab}\Omega \dot{\Psi}\ .
\label{scalcomm4}
\end{equation}

Similar relations hold for the first-order 2-vector $\Psi _{a}$:%
\begin{equation}
\widehat{\dot{\Psi}}_{\bar{a}}-\dot{\widehat{\Psi }}_{\bar{a}}=\left( \Sigma
+\frac{\Theta }{3}\right) \widehat{\Psi }_{\bar{a}}\ ,
\end{equation}%
\begin{equation}
\delta _{a}\dot{\Psi}_{b}-N_{a}^{\,\,\,c}N_{b}^{\,\,\,d}\left( \delta
_{c}\Psi _{d}\right) ^{\cdot }=-\frac{1}{2}\left( \Sigma -\frac{2\Theta }{3}%
\right) \delta _{a}\Psi _{b}\ ,
\end{equation}%
\begin{equation}
\delta _{a}\widehat{\Psi }_{b}-N_{a}^{\,\,\,c}N_{b}^{\,\,\,d}\left( \widehat{%
\delta _{c}\Psi _{d}}\right) =\frac{1}{2}\phi \delta _{a}\Psi _{b}\ ,
\label{scalcomm7}
\end{equation}%
\begin{equation}
\delta _{\lbrack a}\delta _{b]}\Psi _{c}=\frac{1}{2}\mathcal{R}N_{c[a}\Psi
_{b]}\ ,
\end{equation}%
and for the first-order symmetric, trace-free 2-tensor $\Psi _{ab}$:%
\begin{equation}
\widehat{\dot{\Psi}}_{\{ab\}}-\dot{\widehat{\Psi }}_{\{ab\}}=\left( \Sigma +%
\frac{\Theta }{3}\right) \widehat{\Psi }_{\bar{a}}\ ,
\end{equation}%
\begin{equation}
\delta _{a}\dot{\Psi}_{bc}-N_{a}^{\,\,\,d}N_{b}^{\,\,\,e}N_{c}^{\,\,\,f}%
\left( \delta _{d}\Psi _{ef}\right) ^{\cdot }=-\frac{1}{2}\left( \Sigma -%
\frac{2\Theta }{3}\right) \delta _{a}\Psi _{bc}\ ,
\end{equation}%
\begin{equation}
\delta _{a}\widehat{\Psi }_{bc}-N_{a}^{\,\,\,d}N_{b}^{\,\,\,e}N_{c}^{\,\,%
\,f}\left( \widehat{\delta _{d}\Psi _{ef}}\right) =0\ ,
\end{equation}%
\begin{equation}
2\delta _{\lbrack a}\delta _{b]}\Psi _{cd}=\mathcal{R}\left( N_{c[a}\Psi
_{b]d}+N_{d[a}\Psi _{b]c}\right) \ ,
\end{equation}%
where the 2-curvature $\mathcal{R}$ is given by Equation (\ref%
{2TimesGaussianCurv}). In (\ref{scalcomm3}) and (\ref{scalcomm7}) the last
terms on the right-hand sides are second-order in the generic case when $%
\phi =0$ on the background, and can hence be dropped.

\section{Evolution and Propagation Equations and Constraints for $\protect%
\phi=0$}

\label{evolution0}

Here, the evolution and propagation equations and constraints from reference 
\cite{GWKS} for the generic case when $\phi=0$ are repeated. These are
obtained by linearising the 1 + 1 + 2 equations in \cite{1+1+2}.{\footnote{{%
There~are some minor misprints in }\cite{1+1+2}. See \cite{GWKS} for
corrections.}}

The set $S^{(0)}=\{\Theta,\Sigma ,{\mathcal{E}},\mu, p \}$ gives the
quantities which are nonzero on the background. We~define the corresponding
first-order quantities 
\begin{equation}  \label{newvariables0}
\mu_a=\delta_a\mu \, ,\;\; p_a=\delta_a p \, ,\;\; W_a=\delta_a\Theta \,
,\;\; V_a=\delta_a \Sigma \, ,\;\; X_a=\delta_a{\mathcal{E}} \, .
\end{equation}
which vanish on the background. The vorticity-free perturbations are then
given by the following nonzero first-order quantities (which all vanish on
the background): 
\begin{equation}
S^{(1)}\equiv \left\{ X_{a},V_{a},W_{a},\mu _{a},p_{a},{\mathcal{A}},{%
\mathcal{A}}_{a},\Sigma _{a},\Sigma _{ab},{\mathcal{E}}_{a},{\mathcal{E}}%
_{ab},{\mathcal{H}},{\mathcal{H}}_{a},{\mathcal{H}}_{ab},a_{b},\alpha_a,\phi
,\xi ,\zeta _{ab}\right\} ~.
\end{equation}

The evolution equations are given by 
\begin{equation}
\dot{\phi}=\left( \Sigma -\frac{2\Theta }{3}\right) \left( \frac{\phi }{2}-%
\mathcal{A}\right) +\delta ^{a}\alpha _{a}\ ,  \label{phidot0}
\end{equation}%
\begin{equation}
2\dot{\xi}=\left( \Sigma -\frac{2\Theta }{3}\right) \xi +\varepsilon
^{ab}\delta _{a}\alpha _{b}+\mathcal{H}\ ,
\end{equation}%
\begin{equation}
\dot{\mathcal{H}}=\frac{3}{2}\!\!\left( \Sigma -\frac{2\Theta }{3}\right) 
\mathcal{H}-\varepsilon ^{ab}\delta _{a}\mathcal{E}_{b}-3\mathcal{E}\xi \ ,
\end{equation}%
\begin{equation}
\dot{\mu}_{\bar{a}}=\frac{1}{2}\left( \Sigma -\frac{2\Theta }{3}\right) \mu
_{a}-\Theta \left( \mu _{a}+p_{a}\right) -\left( \mu +p\right) W_{a}+\dot{\mu%
}\mathcal{A}_{a}\ ,  \label{odden0}
\end{equation}%
\begin{equation}
\dot{X}_{\bar{a}}=2\!\left( \!\Sigma \!-\!\frac{2\Theta }{3}\!\right)
\!X_{a}\!+\!\frac{3\mathcal{E}}{2}\!\left( \!V_{a}\!\!-\!\frac{2}{3}%
W_{a}\!\right) \!-\!\frac{\mu +p}{2}V_{a}-\frac{\Sigma }{2}\left( \mu
_{a}+p_{a}\right) +\dot{\mathcal{E}}\mathcal{A}_{a}+\varepsilon ^{bc}\delta
_{a}\delta _{b}\mathcal{H}_{c}\ ,
\end{equation}%
\begin{equation}
\dot{V}_{\bar{a}}\!-\!\frac{2}{3}\dot{W}_{\bar{a}}\!=\frac{3}{2}\left(
\Sigma -\frac{2\Theta }{3}\right) \!\left( V_{a}-\frac{2}{3}W_{a}\right)
-X_{a}+\frac{1}{3}\left( \mu _{a}+3p_{a}\right) +\!\left( \!\dot{\Sigma}\!-\!%
\frac{2\dot{\Theta}}{3}\!\right) \!\mathcal{A}_{a}-\delta _{a}\delta ^{b}%
\mathcal{A}_{b},
\end{equation}%
\begin{equation}
\dot{\Sigma}_{\{ab\}}=\left( \Sigma -\frac{2\Theta }{3}\right) \Sigma
_{ab}+\delta _{\{a}\mathcal{A}_{b\}}-\mathcal{E}_{ab}\ ,
\end{equation}%
\begin{equation}
\dot{\zeta}_{\{ab\}}=\frac{1}{2}\left( \Sigma -\frac{2\Theta }{3}\right)
\zeta _{ab}+\delta _{\{a}\alpha _{b\}}-\varepsilon _{c\{a}\mathcal{H}%
_{b\}}^{\,\,\,\,\,c}\ .
\end{equation}

The equations containing both propagation and evolution contributions are%
\begin{equation}
\dot{W}_{\bar{a}}-\delta _{a}\widehat{\mathcal{A}}=\left( \frac{\Sigma }{2}%
-\Theta \right) W_{a}-3\Sigma V_{a}-\frac{1}{2}\left( \mu _{a}+3p_{a}\right)
+\dot{\Theta}\mathcal{A}_{a}+\delta _{a}\delta ^{b}\mathcal{A}_{b}\ ,
\end{equation}%
\begin{equation}
\widehat{\alpha }_{\bar{a}}-\dot{a}_{\bar{a}}=\left( \Sigma +\frac{\Theta }{3%
}\right) \left( \mathcal{A}_{a}+a_{a}\right) -\varepsilon _{ab}\mathcal{H}%
^{b}\ ,
\end{equation}%
\begin{equation}
2\dot{\Sigma}_{\bar{a}}-\widehat{\mathcal{A}}_{a}=\delta _{a}\mathcal{A}%
-\left( \Sigma +\frac{4\Theta }{3}\right) \Sigma _{a}-3\Sigma \alpha _{a}-2%
\mathcal{E}_{a}\ ,
\end{equation}%
\begin{equation}
\dot{\mathcal{E}}_{\bar{a}}\!+\frac{1}{2}\varepsilon _{ab}\widehat{\mathcal{H%
}}^{b}=\!\frac{3}{4}\left( \!\Sigma \!-\frac{4\Theta }{3}\!\right) \!%
\mathcal{E}_{a}+\frac{\left( 3\mathcal{E\!}-\!2\mu \!-\!2p\right) }{4}\Sigma
_{a}+\frac{3}{4}\varepsilon _{ab}\delta ^{b}\mathcal{H\!}\!-\frac{3\mathcal{E%
}}{2}\alpha _{a}\!+\frac{1}{2}\varepsilon _{bc}\delta ^{b}\mathcal{H}%
_{\,\,a}^{c},
\end{equation}%
\begin{equation}
\dot{\mathcal{H}}_{\bar{a}}\!-\frac{1}{2}\varepsilon _{ab}\widehat{\mathcal{E%
}}^{b}=\frac{3}{4}\left( \Sigma -\frac{4\Theta }{3}\right) \mathcal{H}_{a}-%
\frac{3}{4}\varepsilon _{ab}X^{b}-\frac{3\mathcal{E}}{2}\varepsilon _{ab}%
\mathcal{A}^{b}\!+\frac{3\mathcal{E}}{4}\varepsilon _{ab}a^{b}\!-\frac{1}{2}%
\varepsilon _{bc}\delta ^{b}\mathcal{E}_{\,\,a}^{c},
\end{equation}%
\begin{equation}
\dot{\mathcal{E}}_{\{ab\}}\!-\varepsilon _{c\{a}\widehat{\mathcal{H}}%
_{b\}}^{\,\,\,\,\,c}\!=-\frac{3}{2}\!\left( \Sigma \!+\!\frac{2\Theta }{3}%
\right) \!\mathcal{E}_{ab}\!-\!\varepsilon _{c\{a}\delta ^{c}\mathcal{H}%
_{b\}}-\frac{\left( 3\mathcal{E}+\mu +p\right) }{2}\Sigma _{ab},
\end{equation}%
\begin{equation}
\dot{\mathcal{H}}_{\{ab\}}+\varepsilon _{c\{a}\widehat{\mathcal{E}}%
_{b\}}^{\,\,\,\,\,c}=-\frac{3}{2}\left( \Sigma +\frac{2\Theta }{3}\right) 
\mathcal{H}_{ab}+\frac{3\mathcal{E}}{2}\varepsilon _{c\{a}\zeta
_{b\}}^{\,\,\,\,c}+\varepsilon _{c\{a}\delta ^{c}\mathcal{E}_{b\}}\ .
\end{equation}

The pure propagation equations are%
\begin{equation}
\widehat{\phi }=-\left( \Sigma -\frac{2\Theta }{3}\right) \left( \Sigma +%
\frac{\Theta }{3}\right) +\delta ^{a}a_{a}-\mathcal{E}-\frac{2\left( \mu
+\Lambda \right) }{3},
\end{equation}%
\begin{equation}
2\widehat{\xi }=\varepsilon ^{ab}\delta _{a}a_{b}\ ,
\end{equation}%
\begin{equation}
\widehat{\mathcal{H}}=-\delta ^{a}\mathcal{H}_{a}\ ,
\end{equation}%
\begin{equation}
\widehat{\mathcal{A}}_{a}=\delta _{a}\mathcal{A}\ ,
\end{equation}%
\begin{equation}
\widehat{p}_{\bar{a}}=-\left( \mu +p\right) \delta _{a}\mathcal{A}\ ,
\end{equation}%
\begin{equation}
\frac{\widehat{\mu }_{\bar{a}}}{3}-\widehat{X}_{\bar{a}}=\frac{3\mathcal{E}}{%
2}\delta _{a}\phi +\delta _{a}\delta ^{b}\mathcal{E}_{b}\!\!\ ,
\end{equation}%
\begin{equation}
\frac{2}{3}\widehat{W}_{\bar{a}}-\widehat{V}_{\bar{a}}=\frac{3\Sigma }{2}%
\delta _{a}\phi +\delta _{a}\delta ^{b}\Sigma _{b}\ ,
\end{equation}%
\begin{equation}
\widehat{\Sigma }_{\bar{a}}=\frac{1}{2}\left( V_{a}+\frac{4}{3}W_{a}\right) -%
\frac{3\Sigma }{2}a_{a}-\delta ^{b}\Sigma _{ab}\ ,
\end{equation}%
\begin{equation}
2\widehat{\mathcal{E}}_{\bar{a}}=X_{a}-3\mathcal{E}a_{a}-3\Sigma \varepsilon
_{ab}\mathcal{H}^{b}-2\delta ^{b}\mathcal{E}_{ab}\!+\frac{2}{3}\mu _{a}\ ,
\end{equation}%
\begin{equation}
2\widehat{\mathcal{H}}_{\bar{a}}=\delta _{a}\mathcal{H}-3\mathcal{E}%
\varepsilon _{ab}\Sigma ^{b}+3\Sigma \varepsilon _{ab}\mathcal{E}%
^{b}-2\delta ^{b}\mathcal{H}_{ab}\ ,
\end{equation}%
\begin{equation}
\widehat{\Sigma }_{\{ab\}}=\delta _{\{a}\Sigma _{b\}}+\frac{3\Sigma }{2}%
\zeta _{ab}-\varepsilon _{c\{a}\mathcal{H}_{b\}}^{\,\,\,\,\,c}\ ,
\end{equation}%
\begin{equation}
\widehat{\zeta }_{\{ab\}}=\left( \Sigma +\frac{\Theta }{3}\right) \Sigma
_{ab}+\delta _{\{a}a_{b\}}-\mathcal{E}_{ab}\ .
\end{equation}

Finally, the constraints are%
\begin{equation}
\varepsilon ^{ab}\delta _{a}\mathcal{A}_{b}=0\ ,  \label{Abconstr0}
\end{equation}%
\begin{equation}
p_{a}=-\left( \mu +p\right) \mathcal{A}_{a}\ ,
\end{equation}%
\begin{equation}
\varepsilon ^{ab}\delta _{a}\Sigma _{b}=-3\Sigma \xi +\mathcal{H}\ ,
\end{equation}%
\begin{equation}
\varepsilon _{ab}\delta ^{b}{\xi}+\!\!\delta ^{b}\zeta _{ab}\!-\!\frac{%
\delta _{a}\phi }{2}=\frac{1}{2}\left( \!\Sigma \!-\frac{2\Theta }{3}%
\!\right) \!\Sigma _{a}\!+\mathcal{E}_{a},
\end{equation}%
\begin{equation}
\left( V_{a}-\frac{2}{3}W_{a}\right) +2\delta ^{b}\Sigma _{ab}=-2\varepsilon
_{ab}\mathcal{H}^{b}\ .  \label{constr30}
\end{equation}

\section{Evolution and Propagation Equations and Constraints for $\protect%
\phi\neq 0$}

\label{evolution}

Here, the evolution and propagation equations and constraints are given for
the exceptional case where $\phi \neq 0$ on the background, corresponding to
the negatively curved Friedmann universe. The set $S^{(0)}=\{\Theta ,\mu
,p,\phi \}$ gives the quantities which are nonzero on the background. We
define corresponding first-order quantities 
\begin{equation}
\mu _{a}=\delta _{a}\mu \,,\;\;p_{a}=\delta _{a}p\,,\;\;W_{a}=\delta
_{a}\Theta \,,\;\;\phi _{a}=\delta _{a}\phi \,,
\end{equation}%
which vanish on the background. As before, we let $V_{a}=\delta _{a}\Sigma $
and $X_{a}=\delta _{a}{\mathcal{E}}$. Vorticity-free perturbations hence are
given by the following nonzero first-order quantities (which all vanish on
the background): 
\begin{equation}
S^{(1)}\equiv \left\{ \phi _{a},W_{a},\mu _{a},p_{a},{\mathcal{A}},{\mathcal{%
A}}_{a},\Sigma ,\Sigma _{a},\Sigma _{ab},{\mathcal{E}},{\mathcal{E}}_{a},{%
\mathcal{E}}_{ab},{\mathcal{H}},{\mathcal{H}}_{a},{\mathcal{H}}%
_{ab},a_{b},\alpha _{a},\xi ,\zeta _{ab}\right\} ~.
\end{equation}

The following evolution equations then hold on the perturbed spacetimes: 
\begin{equation}
\dot{\phi}_{\bar{a}}=\dot{\phi}{\mathcal{A}}_{a}-\frac{2\Theta }{3}\phi
_{a}+\left( V_{a}-\frac{2}{3}W_{a}\right) \frac{\phi }{2}+\frac{2\Theta }{3}%
\delta _{a}\mathcal{A}+\delta _{a}\delta ^{b}\alpha _{b}\ ,  \label{phidot}
\end{equation}%
\begin{equation}
2\dot{\xi}=-\frac{2\Theta }{3}\xi +\varepsilon ^{ab}\delta _{a}\alpha _{b}+%
\mathcal{H}\ ,
\end{equation}%
\begin{equation}
\dot{\mathcal{H}}=-\Theta \mathcal{H}-\varepsilon ^{ab}\delta _{a}\mathcal{E}%
_{b}\ ,
\end{equation}%
\begin{equation}
\dot{\mu}_{\bar{a}}=-\frac{\Theta }{3}\mu _{a}-\Theta \left( \mu
_{a}+p_{a}\right) -\left( \mu +p\right) W_{a}+\dot{\mu}\mathcal{A}_{a}\ ,
\label{odden}
\end{equation}%
\begin{equation}
\dot{\mathcal{\ E}}=-\Theta {\mathcal{E}}-\frac{1}{2}(\mu +p)\Sigma
+\varepsilon ^{ab}\delta _{a}\mathcal{H}_{b}\ ,
\end{equation}%
\begin{equation}
\dot{\Sigma}_{\{ab\}}=-\frac{2\Theta }{3}\Sigma _{ab}+\delta _{\{a}\mathcal{A%
}_{b\}}-\mathcal{E}_{ab}\ ,
\end{equation}%
\begin{equation}
\dot{\zeta}_{\{ab\}}=-\frac{\Theta }{3}\zeta _{ab}+\delta _{\{a}\alpha
_{b\}}-\varepsilon _{c\{a}\mathcal{H}_{b\}}^{\,\,\,\,\,c}\ .
\end{equation}

The equations containing both propagation and evolution contributions are%
\begin{equation}
\dot{\Sigma}-\frac{2}{3}\widehat{\mathcal{A}}=-\frac{1}{3}\phi {\mathcal{A}}-%
\frac{2}{3}\Theta \Sigma -\frac{1}{3}\delta _{a}{\mathcal{A}}^{a}-{\mathcal{E%
}}\,,
\end{equation}%
\begin{equation}
\dot{W}_{\bar{a}}-\delta _{a}\widehat{\mathcal{A}}=-\Theta W_{a}-\frac{1}{2}%
\left( \mu _{a}+3p_{a}\right) +\dot{\Theta}\mathcal{A}_{a}+\delta _{a}\delta
^{b}\mathcal{A}_{b}+\phi \delta _{a}{\mathcal{A}}\ ,
\end{equation}%
\begin{equation}
\widehat{\alpha }_{\bar{a}}-\dot{a}_{\bar{a}}=\frac{\Theta }{3}\left( 
\mathcal{A}_{a}+a_{a}\right) -\varepsilon _{ab}\mathcal{H}-\frac{1}{2}\phi
\alpha _{a}\ ,
\end{equation}%
\begin{equation}
2\dot{\Sigma}_{\bar{a}}-\widehat{\mathcal{A}}_{a}=\delta _{a}\mathcal{A}-%
\frac{4\Theta }{3}\Sigma _{a}-2\mathcal{E}_{a}-\frac{1}{2}\phi {\mathcal{A}}%
_{a}\ ,
\end{equation}%
\begin{equation}
\dot{\mathcal{E}}_{\bar{a}}\!+\frac{1}{2}\varepsilon _{ab}\widehat{\mathcal{H%
}}^{b}=-\Theta \,\mathcal{E}_{a}-\frac{1}{2}\left( \mu +p\right) \Sigma _{a}+%
\frac{3}{4}\varepsilon _{ab}\delta ^{b}\mathcal{H\!}\!+\frac{1}{2}%
\varepsilon _{bc}\delta ^{b}\mathcal{H}_{\,\,a}^{c}-\frac{1}{4}\phi
\varepsilon _{ab}{\mathcal{H}}^{b},
\end{equation}%
\begin{equation}
\dot{\mathcal{H}}_{\bar{a}}\!-\frac{1}{2}\varepsilon _{ab}\widehat{\mathcal{E%
}}^{b}=-\Theta \mathcal{H}_{a}-\frac{3}{4}\varepsilon _{ab}X^{b}-\frac{1}{2}%
\varepsilon _{bc}\delta ^{b}\mathcal{E}_{\,\,a}^{c}+\frac{1}{4}\phi
\varepsilon _{ab}{\mathcal{E}}^{b},
\end{equation}%
\begin{equation}
\dot{\mathcal{E}}_{\{ab\}}\!-\varepsilon _{c\{a}\widehat{\mathcal{H}}%
_{b\}}^{\,\,\,\,\,c}\!=-\Theta \,\mathcal{E}_{ab}\!-\!\varepsilon
_{c\{a}\delta ^{c}\mathcal{H}_{b\}}-\frac{\left( \mu +p\right) }{2}\Sigma
_{ab}+\frac{1}{2}\phi \varepsilon _{c\{a}{\mathcal{H}}_{b\}}^{c},
\end{equation}%
\begin{equation}
\dot{\mathcal{H}}_{\{ab\}}+\varepsilon _{c\{a}\widehat{\mathcal{E}}%
_{b\}}^{\,\,\,\,\,c}=-\Theta \,\mathcal{H}_{ab}+\varepsilon _{c\{a}\delta
^{c}\mathcal{E}_{b\}}-\frac{1}{2}\phi \varepsilon _{c\{a}{\mathcal{E}}%
_{b\}}^{c}\ .
\end{equation}

The pure propagation equations are%
\begin{equation}
\widehat{\phi }_{\bar{a}}=\frac{1}{3}\left( V_{a}+\frac{4}{3}W_{a}\right)
\Theta +\delta _{a}\delta ^{b}a_{b}-X_{a}-\frac{2}{3}\mu _{a}-\frac{3}{2}%
\phi \phi _{a}~,
\end{equation}%
\begin{equation}
2\widehat{\xi }=\varepsilon ^{ab}\delta _{a}a_{b}-2\phi \xi \ ,
\end{equation}%
\begin{equation}
\widehat{\mathcal{H}}=-\delta ^{a}\mathcal{H}_{a}-\frac{3}{2}\phi {\mathcal{H%
}}\ ,
\end{equation}%
\begin{equation}
\widehat{\mathcal{A}}_{a}=\delta _{a}\mathcal{A}-\frac{1}{2}\phi {\mathcal{A}%
}_{a}\ ,
\end{equation}%
\begin{equation}
\widehat{p}_{\bar{a}}=-\left( \mu +p\right) \delta _{a}\mathcal{A}-\frac{1}{2%
}\phi p\ ,
\end{equation}%
\begin{equation}
\frac{\widehat{\mu }_{\bar{a}}}{3}-\widehat{X}_{\bar{a}}=\delta _{a}\delta
^{b}\mathcal{E}_{b}+2\phi X_{a}-\frac{1}{6}\phi \mu _{a}\!\!\ ,
\end{equation}%
\begin{equation}
\frac{2}{3}\widehat{W}_{\bar{a}}-\widehat{V}_{\bar{a}}=\frac{3\Sigma }{2}%
\delta _{a}\phi +\delta _{a}\delta ^{b}\Sigma _{b}+2\phi V_{a}-\frac{1}{3}%
\phi W_{a}\ ,
\end{equation}%
\begin{equation}
\widehat{\Sigma }_{\bar{a}}=\frac{1}{2}\left( V_{a}+\frac{4}{3}W_{a}\right)
-\delta ^{b}\Sigma _{ab}-\frac{3}{2}\phi \Sigma _{ab}\ ,
\end{equation}%
\begin{equation}
2\widehat{\mathcal{E}}_{\bar{a}}=X_{a}-2\delta ^{b}\mathcal{E}_{ab}\!+\frac{2%
}{3}\mu _{a}-3\phi {\mathcal{E}}_{a}\ ,
\end{equation}%
\begin{equation}
2\widehat{\mathcal{H}}_{\bar{a}}=\delta _{a}\mathcal{H}-2\delta ^{b}\mathcal{%
H}_{ab}-3\phi {\mathcal{H}}_{a}\ ,
\end{equation}%
\begin{equation}
\widehat{\Sigma }_{\{ab\}}=\delta _{\{a}\Sigma _{b\}}-\varepsilon _{c\{a}%
\mathcal{H}_{b\}}^{\,\,\,\,\,c}-\frac{1}{2}\phi \Sigma _{ab}\ ,
\end{equation}%
\begin{equation}
\widehat{\zeta }_{\{ab\}}=\frac{\Theta }{3}\Sigma _{ab}+\delta _{\{a}a_{b\}}-%
\mathcal{E}_{ab}-\phi \zeta _{ab}\ .
\end{equation}

Finally, the constraints are%
\begin{equation}
\varepsilon ^{ab}\delta _{a}\mathcal{A}_{b}=0\ ,  \label{Abconstr}
\end{equation}%
\begin{equation}
p_{a}=-\left( \mu +p\right) \mathcal{A}_{a}\ ,
\end{equation}%
\begin{equation}
\varepsilon ^{ab}\delta _{a}\Sigma _{b}=\mathcal{H}\ ,
\end{equation}%
\begin{equation}
\varepsilon _{ab}\delta ^{b}\xi +\!\!\delta ^{b}\zeta _{ab}\!-\!\frac{\phi
_{a}}{2}=-\frac{\Theta }{3}\,\Sigma _{a}\!+\mathcal{E}_{a},
\end{equation}%
\begin{equation}
\left( V_{a}-\frac{2}{3}W_{a}\right) +2\delta ^{b}\Sigma _{ab}=-2\varepsilon
_{ab}\mathcal{H}^{b}-\phi \Sigma _{a}\ .  \label{constr3}
\end{equation}

\section{Harmonics}

\label{harmonics}

In this Appendix we enlist a set of identities for the even $Q_{a}^{k_{\perp
}}$ and odd $\overline{Q}_{a}^{k_{\perp }}$ vector harmonics, including the
orthogonality relations%
\begin{equation}
N^{ab}Q_{a}^{k_{\perp }}\overline{Q}_{b}^{k_{\perp }}=0~,
\end{equation}%
the algebraic relations%
\begin{equation}
Q_{a}^{k_{\perp }}=-\varepsilon _{a}^{\,\,\,\,b}\overline{Q}_{b}^{k_{\perp
}}\ ,\quad \overline{Q}_{a}^{k_{\perp }}=\varepsilon
_{a}^{\,\,\,\,b}Q_{b}^{k_{\perp }}\ ,\ \ 
\end{equation}%
and the differential relations%
\begin{equation}
\dot{Q}_{a}^{k_{\perp }}=\widehat{Q}_{a}^{k_{\perp }}=0\ ,\quad \dot{%
\overline{Q}}_{a}^{k_{\perp }}=\widehat{\overline{Q}}_{a}^{k_{\perp }}=0\ ,
\end{equation}%
\begin{equation}
\delta ^{2}Q_{a}^{k_{\perp }}=\frac{{\mathcal{R}}a_{2}^{2}-2k_{\perp }^{2}}{%
2a_{2}^{2}}Q_{a}^{k_{\perp }}\ ,\quad \delta ^{2}\overline{Q}_{a}^{k_{\perp
}}=\frac{{\mathcal{R}}a_{2}^{2}-2k_{\perp }^{2}}{2a_{2}^{2}}\overline{Q}%
_{a}^{k_{\perp }}\ ,
\end{equation}%
\begin{equation}
\delta ^{a}Q_{a}^{k_{\perp }}=-\frac{k_{\perp }^{2}}{a_{2}}Q^{k_{\perp }}\
,\quad \delta ^{a}\overline{Q}_{a}^{k_{\perp }}=0\ ,  \label{vectspherid3}
\end{equation}%
\begin{equation}
\varepsilon ^{ab}\delta _{a}Q_{b}^{k_{\perp }}=0\ ,\quad \varepsilon
^{ab}\delta _{a}\overline{Q}_{b}^{k_{\perp }}=\frac{k_{\perp }^{2}}{a_{2}}%
Q^{k_{\perp }}~,  \label{vectspherid4}
\end{equation}%
where the 2-curvature $\mathcal{R}$ is given by Equation (\ref%
{2TimesGaussianCurv}).

The even and odd tensor spherical harmonics obey the orthogonality relations%
\begin{equation}
N^{ab}N^{cd}Q_{ac}^{k_{\perp }}\overline{Q}_{bd}^{k_{\perp }}=0~,
\end{equation}%
the algebraic relations%
\begin{equation}
Q_{ab}^{k_{\perp }}=\varepsilon _{\{a}^{\,\,\,\,\,\,\,c}\overline{Q}%
_{b\}c}^{k_{\perp }}\ ,\quad \overline{Q}_{ab}^{k_{\perp }}=-\varepsilon
_{\{a}^{\,\,\,\,\,\,\,c}Q_{b\}c}^{k_{\perp }}\ ,\ \ 
\end{equation}%
and the differential relations%
\begin{equation}
\dot{Q}_{ab}^{k_{\perp }}=\widehat{Q}_{ab}^{k_{\perp }}=0\ ,\quad \dot{%
\overline{Q}}_{ab}^{k_{\perp }}=\widehat{\overline{Q}}_{ab}^{k_{\perp }}=0\ ,
\end{equation}%
\begin{equation}
\delta ^{2}Q_{ab}^{k_{\perp }}=\frac{2{\mathcal{R}}a_2^2-k_{\perp }^{2}}{%
a_{2}^{2}}Q_{ab}^{k_{\perp }}\ ,\quad \delta ^{2}\overline{Q}_{ab}^{k_{\perp
}}= \frac{2{\mathcal{R}}a_2^2-k_{\perp }^{2}}{a_{2}^{2}}\overline{Q}%
_{ab}^{k_{\perp }}~,
\end{equation}%
\begin{equation}
\delta ^{b}Q_{ab}^{k_{\perp }}=\frac{{\mathcal{R}}a_2^2-k_{\perp }^{2}}{%
a_{2}^{2}}Q_{a}^{k_{\perp }}\ ,\quad \delta ^{b}\overline{Q}_{ab}^{k_{\perp
}}= -\frac{{\mathcal{R}}a_2^2-k_{\perp }^{2}}{a_{2}^{2}}\overline{Q}%
_{a}^{k_{\perp }}~,
\end{equation}%
\begin{equation}
\varepsilon _{a}^{\,\,\,\,\,c}\delta ^{b}Q_{bc}^{k_{\perp }}= \frac{{%
\mathcal{R}}a_2^2-k_{\perp }^{2}}{a_{2}^{2}}\overline{Q}_{a}^{k_{\perp }}\
,\quad \varepsilon _{a}^{\,\,\,\,\,c}\delta ^{b}\overline{Q}_{bc}^{k_{\perp
}}=\frac{{\mathcal{R}}a_2^2-k_{\perp }^{2}}{a_{2}^{2}}Q_{a}^{k_{\perp }}~,
\end{equation}%
\begin{equation}
\varepsilon ^{bc}\delta _{b}Q_{ac}^{k_{\perp }}=\frac{{\mathcal{R}}%
a_2^2-k_{\perp }^{2}}{a_{2}^{2}}\overline{Q}_{a}^{k_{\perp }}\ ,\quad
\varepsilon ^{bc}\delta _{b}\overline{Q}_{ac}^{k_{\perp }}=\frac{{\mathcal{R}%
}a_2^2-k_{\perp }^{2}}{a_{2}^{2}}Q_{a}^{k_{\perp }}~.
\end{equation}

\section{Harmonic Coefficients}

\label{harmoniccoefficients} The perturbation variables coupled to $\mu
_{k_{\parallel }k_{\perp }}^{V}$, $\Sigma _{k_{\parallel }k_{\perp }}^{T} $, 
$\mathcal{E}_{k_{\parallel }k_{\perp }}^{T}$, and $\overline{\mathcal{H}}%
_{k_{\parallel }k_{\perp }}^{T}$ are 
\begin{equation}
{\mathcal{A}}_{k_{\parallel }k_{\perp }}^{V}=\frac{a_{1}}{ik_{\parallel }}%
\frac{{\mathcal{A}}_{k_{\parallel }k_{\perp }}^{S}}{a_{2}}=-\frac{c_{s}^{2}}{%
\mu +p}\mu _{k_{\parallel }k_{\perp }}^{V}~,  \label{constrE}
\end{equation}%
\begin{equation}
\frac{ik_{\parallel }}{a_{1}}\zeta _{k_{\parallel }k_{\perp }}^{T}=\left(
\Sigma +\frac{\Theta }{3}\right) \Sigma _{k_{\parallel }k_{\perp }}^{T}-{%
\mathcal{E}}_{k_{\parallel }k_{\perp }}^{T}~,
\end{equation}%
\begin{equation}
\frac{ik_{\parallel }}{a_{1}a_{2}}\Sigma _{k_{\parallel }k_{\perp
}}^{V}=-\left( B+\frac{{\mathcal{R}}a_2^2-k_{\perp }^{2}}{a_{2}^{2}}\right) 
\frac{\Sigma _{k_{\parallel }k_{\perp }}^{T}}{2}+\frac{3}{2}\Sigma {\mathcal{%
E}}_{k_{\parallel }k_{\perp }}^{T}-i\frac{k_{\parallel }}{a_{1}}\overline{%
\mathcal{H}}_{k_{\parallel }k_{\perp }}^{T}~,
\end{equation}%
\begin{equation}
\frac{2\overline{\mathcal{H}}_{k_{\parallel }k_{\perp }}^{V}}{3a_{2}}=-\frac{%
\Sigma }{a_{2}B}\mu _{k_{\parallel }k_{\perp }}^{V}+\Sigma C{\mathcal{E}}%
_{k_{\parallel }k_{\perp }}^{T}-{\mathcal{E}}\Sigma _{k_{\parallel }k_{\perp
}}^{T}-\frac{ik_{\parallel }}{a_{1}}J\frac{\overline{\mathcal{H}}%
_{k_{\parallel }k_{\perp }}^{T}}{3}~,  \label{HbarVconstr}
\end{equation}%
\begin{equation}
\frac{V_{k_{\parallel }k_{\perp }}^{V}}{a_{2}}=-\frac{2\Sigma }{a_{2}B}\mu
_{k_{\parallel }k_{\perp }}^{V}-\frac{B}{3}\left( 1+2C\right) \Sigma
_{k_{\parallel }k_{\perp }}^{T}+\Sigma \left( 1+2C\right) \mathcal{E}%
_{k_{\parallel }k_{\perp }}^{T}-\frac{2}{3}\frac{ik_{\parallel }}{a_{1}}%
\left( 1+J\right) \overline{\mathcal{H}}_{k_{\parallel }k_{\perp }}^{T}~,
\label{VVconstr}
\end{equation}%
\begin{equation}
\begin{array}{rlll}
\label{WVconstr} \frac{W_{k_{\parallel }k_{\perp }}^{V}}{a_{2}} & = & \frac{%
3\Sigma }{2a_{2}B}\mu _{k_{\parallel }k_{\perp }}^{V}+\left( 3{\mathcal{E}}+%
\frac{{\mathcal{R}}a_2^2-k_{\perp }^{2}}{a_{2}^{2}}-B\right) \frac{\Sigma
_{k_{\parallel }k_{\perp }}^{T}}{2} &  \\ 
&  & \!\!\!\!\!\!\!\!\!\!+\frac{3\Sigma }{2}\left( 1-C\right) \mathcal{E}%
_{k_{\parallel }k_{\perp }}^{T}-\frac{ik_{\parallel }}{2a_{1}}\left(
2-J\right) \overline{\mathcal{H}}_{k_{\parallel }k_{\perp }}^{T}~, &  \\ 
&  &  & 
\end{array}%
\end{equation}%
\begin{equation}
\frac{ik_{\parallel }}{a_{1}}\frac{\phi _{k_{\parallel }k_{\perp }}^{S}}{%
a_{2}^{2}}=-\frac{k_{\parallel }^{2}}{a_{1}^{2}}\frac{2}{a_{2}B}\mu
_{k_{\parallel }k_{\perp }}^{V}-\frac{BL}{3 }\Sigma _{k_{\parallel }k_{\perp
}}^{T}+\left( L\Sigma-\frac{{\mathcal{R}}a_2^2-k_{\perp }^{2}}{a_{2}^{2}B}%
\frac{k_{\perp }^{2}}{a_{2}^{2}}\right) \mathcal{E}_{k_{\parallel }k_{\perp
}}^{T}-\frac{ik_{\parallel }}{a_{1}}\frac{2L}{3 }\overline{\mathcal{H}}%
_{k_{\parallel }k_{\perp }}^{T}~,
\end{equation}%
\begin{equation}
\begin{array}{rlll}
\frac{ik_{\parallel }}{a_{1}a_{2}}\alpha _{k_{\parallel }k_{\perp }}^{V} & =
& -\left[ \frac{3\Sigma }{2B}+\left( \Sigma +\frac{\Theta }{3}\right) \frac{%
c_{s}^{2}}{\mu +p}\right] \frac{\mu _{k_{\parallel }k_{\perp }}^{V}}{a_{2}}
&  \\ 
&  & -\frac{3\mathcal{E}}{2}\Sigma _{k_{\parallel }k_{\perp }}^{T}+\frac{%
3\Sigma }{2}C\mathcal{E}_{k_{\parallel }k_{\perp }}^{T}-\frac{ik_{\parallel }%
}{2a_{1}}J\overline{\mathcal{H}}_{k_{\parallel }k_{\perp }}^{T}~, &  \\ 
&  &  & 
\end{array}%
\end{equation}%
\begin{equation}
\begin{array}{rlll}
\label{XVconstr} \frac{X_{k_{\parallel }k_{\perp }}^{V}}{a_{2}} & = & \left[
1-\frac{3}{B}\left( \frac{k_{\perp }^{2}}{a_{2}^{2}}+3\mathcal{E}\right) %
\right] \frac{\mu _{k_{\parallel }k_{\perp }}^{V}}{3a_{2}}-{\mathcal{E}}%
\left( \Theta-\frac{3}{2}\Sigma\right) \Sigma _{k_{\parallel }k_{\perp
}}^{T}+C\left( \frac{k_{\perp }^{2}}{a_{2}^{2}}+3\mathcal{E}\right) \mathcal{%
E}_{k_{\parallel }k_{\perp }}^{T} &  \\ 
&  & +\frac{a_{1}}{ik_{\parallel }}\left[ \mathcal{E}\left( \frac{1}{3}%
\Theta-\frac{1}{2}\Sigma\right) \frac{k_{\parallel }^{2}}{a_{1}^{2}}-\Sigma%
\frac{{\mathcal{R}}a_2^2-k_{\perp }^{2}}{4a_{2}^{2}}\frac{k_{\perp }^{2}}{%
a_{2}^{2}}\right] \frac{6\overline{\mathcal{H}}_{k_{\parallel }k_{\perp
}}^{T}}{ B}~, & 
\end{array}%
\end{equation}%
\begin{equation}
\begin{array}{rlll}
\frac{ik_{\parallel }}{a_{1}a_{2}}\mathcal{E}_{k_{\parallel }k_{\perp
}}^{V}\! & = & \frac{k_{\parallel }^{2}}{a_{1}^{2}}\frac{\mu _{k_{\parallel
}k_{\perp }}^{V}}{a_{2}B}\!\!-\frac{3\mathcal{E}}{2}\left( \Sigma +\frac{%
\Theta }{3}\right) \Sigma _{k_{\parallel }k_{\perp }}^{T}+\left( \frac{3%
\mathcal{E}}{2}-C\frac{k_{\parallel }^{2}}{a_{1}^{2}}\right) \mathcal{E}%
_{k_{\parallel }k_{\perp }}^{T} &  \\ 
&  & -\frac{ik_{\parallel }}{a_{1}}\frac{3\overline{\mathcal{H}}%
_{k_{\parallel }k_{\perp }}^{T}}{2B}\left[ 2\mathcal{E}\left( \Sigma +\frac{%
\Theta }{3}\right) +\Sigma \frac{{\mathcal{R}}a_2^2-k_{\perp }^{2}}{a_{2}^{2}%
}\right] ~, &  \\ 
&  &  & 
\end{array}%
\end{equation}%
and which are coupled to $\overline{\mathcal{E}}_{k_{\parallel }k_{\perp
}}^{T}$, $\mathcal{H}_{k_{\parallel }k_{\perp }}^{T}$ are%
\begin{equation}
\begin{array}{rlll}
{\mathcal{H}}_{k_{\parallel }k_{\perp }}^{V} & = & \frac{ik_{\parallel }}{%
a_{1}}\frac{a_{2}}{k_{\perp }^{2}}{\mathcal{H}}_{k_{\parallel }k_{\perp
}}^{S}=-\frac{ik_{\parallel }}{a_{1}}\overline{\alpha }_{k_{\parallel
}k_{\perp }}^{V}=\frac{ik_{\parallel }}{a_{1}}\overline{\Sigma }%
_{k_{\parallel }k_{\perp }}^{V} &  \\ 
& = & \frac{{\mathcal{R}}a_2^2-k_{\perp }^{2}}{2a_{2}}\overline{\Sigma }%
_{k_{\parallel }k_{\perp }}^{T}=\frac{{\mathcal{R}}a_2^2-k_{\perp }^{2}}{%
a_{2}B}\left( \frac{3\Sigma }{2}\overline{\mathcal{E}}_{k_{\parallel
}k_{\perp }}^{T}+\frac{ik_{\parallel }}{a_{1}}{\mathcal{H}}_{k_{\parallel
}k_{\perp }}^{T}\right) ~, &  \\ 
&  &  & 
\end{array}%
\end{equation}%
\begin{equation}
\frac{B}{2}\overline{\zeta }_{k_{\parallel }k_{\perp }}^{T}=\left( \Sigma +%
\frac{\Theta }{3}\right) {\mathcal{H}}_{k_{\parallel }k_{\perp }}^{T}-\frac{%
a_{1}}{ik_{\parallel }}\left( \frac{k_{\parallel }^{2}}{a_{1}^{2}}-\frac{{%
\mathcal{R}}a_2^2-k_{\perp }^{2}}{2a_{2}^{2}}\right) \overline{\mathcal{E}}%
_{k_{\parallel }k_{\perp }}^{T}~,
\end{equation}%
\begin{equation}
\overline{\mathcal{E}}_{k_{\parallel }k_{\perp }}^{V}=\frac{{\mathcal{R}}%
a_2^2-k_{\perp }^{2}}{2a_{2}}\left[ \frac{a_{1}}{ik_{\parallel }}\left( 1-%
\frac{9\Sigma ^{2}}{2B}\right) \overline{\mathcal{E}}_{k_{\parallel
}k_{\perp }}^{T}-\frac{3\Sigma }{B}\mathcal{H}_{k_{\parallel }k_{\perp }}^{T}%
\right] ~,  \label{constrF}
\end{equation}
where

\begin{equation}
B\equiv \frac{2k_{\parallel }^{2}}{a_{1}^{2}}+\frac{k_{\perp }^{2}}{a_{2}^{2}%
}+\frac{9\Sigma ^{2}}{2}+3\mathcal{E}=\frac{2k_{\parallel }^{2}}{a_{1}^{2}}-%
\frac{{\mathcal{R}}a_2^2-k_{\perp }^{2}}{a_{2}^{2}}+3\Sigma \left( \Sigma +%
\frac{\Theta }{3}\right) \ ,  \label{Bdef0}
\end{equation}%
\begin{equation}
CB\equiv\left( \frac{{\mathcal{R}}a_2^2-k_{\perp }^{2}}{a_{2}^{2}}+3\mathcal{%
E}\right)= \Sigma\left(\Theta-\frac{3}{2}\Sigma\right)-\frac{k_\perp^2}{a_2^2%
} \ ,  \label{Cdef0}
\end{equation}%
\begin{equation}
LB\equiv \frac{1}{\Sigma}\left(2CB\frac{k_{\parallel }^{2}}{a_{1}^{2}}+\frac{%
k_{\perp }^{2}}{a_{2}^{2}}\left( B+\frac{{\mathcal{R}}a_2^2-k_{\perp }^{2}}{%
a_{2}^{2}}\right)\right)= 3\Sigma\left(\frac{k_{\perp }^{2}}{a_{2}^{2}}-%
\frac{k_{\parallel }^{2}}{a_{1}^{2}}\right) +\Theta\left(\frac{2k_{\parallel
}^{2}}{a_{1}^{2}}+\frac{k_{\perp }^{2}}{a_{2}^{2}}\right)\, ,
\end{equation}%
\begin{equation}
\begin{array}{rlll}
JB\equiv \frac{({\mathcal{R}}a_2^2-k_{\perp }^{2})k_{\perp }^{2}a_{1}^{2}}{%
k_{\parallel }^{2}a_{2}^{4}}+2CB=\left({\mathcal{R}}-\frac{k_{\perp }^{2}}{%
a_{2}^{2}}\right) \left(\frac{2k_{\parallel }^{2}}{a_{1}^{2}}+\frac{k_{\perp
}^{2}}{a_{2}^{2}}\right)\frac{a_1^2}{k_{\parallel }^{2}} +6{\mathcal{E}} & 
&  &  \\ 
= \frac{k_\perp^2 a_1^2}{k_\parallel ^2 a_2^2}\left({\mathcal{R}}-\frac{%
2k_{\parallel }^{2}}{a_{1}^{2}}-\frac{k_{\perp }^{2}}{a_{2}^{2}}
\right)+2\Sigma\left(\Theta-\frac{3}{2}\Sigma\right)\, . &  &  &  \\ 
&  &  & 
\end{array}%
\end{equation}

\section{Shear and Weyl Tensor Tensorial Waves in Friedmann Spacetime}

\label{App3dim}

In Friedmann spacetimes the background equations [see Equations (3.29),
(3.34) and (3.40) of~\cite{Bonometto}, respectively, specified for Friedmann
spacetimes] are:

\begin{equation}
\dot{\mu}+\Theta \left( \mu +p\right) =0~,  \label{drho}
\end{equation}%
\begin{equation}
\dot{\Theta}+\frac{\Theta ^{2}}{3}+\frac{\mu +3p}{2}-\Lambda =0~,
\label{dTheta}
\end{equation}%
\begin{equation}
\frac{\Theta ^{2}}{3}+\frac{3K}{a^{2}}=\mu +\Lambda ~,  \label{Theta}
\end{equation}%
with curvature $6K/a^{2}=~^{\left( 3\right) }R$ of 3-spaces of homogeneity,
where $^{\left( 3\right) }R$ is the three-dimensional Ricci scalar.

In Friedmann spacetimes the perturbations are classified into scalar,
vector, and tensor types. 
The~pure tensor perturbations are characterized by vanishing of vorticity
and of all gauge-invariant vectors and scalars at the first order \cite%
{DunsbyBassettEllis,perturb4}. Thus, the acceleration and the gradient of
all scalars also vanish. Moreover we assume the perturbed energy momentum
tensor to describe a barotropic perfect~fluid.

The evolution and constraint equations for the shear and for the electric
and magnetic parts of Weyl tensor in 1 + 3 covariant formalism [see
Equations (7)--(9), (12), (13) and (21),of~\cite{DunsbyBassettEllis}%
,~respectively]~are:

\begin{equation}
\dot{\sigma}_{<ab>}+\frac{2}{3}\Theta \sigma _{ab}+E_{ab}=0~,
\label{sigmaEq}
\end{equation}%
\begin{equation}
\dot{E}_{<ab>}+\Theta E_{ab}-\mathrm{curl} H_{ab}+\frac{\mu +p}{2}\sigma
_{ab}=0~,  \label{Eeq}
\end{equation}%
\begin{equation}
\dot{H}_{<ab>}+\Theta H_{ab}+\mathrm{curl}E_{ab}=0~,  \label{HabEq}
\end{equation}%
\begin{equation}
H_{ab}=\mathrm{curl}\sigma _{ab}~,  \label{Heq}
\end{equation}%
\begin{equation}
D^{a}H_{ab}=0~,  \label{divfreeH}
\end{equation}%
\begin{equation}
D^{a}E_{ab}=0~,  \label{divfreeE}
\end{equation}%
where the $\mathrm{curl}$ of a tensor $T_{ab}$ is given by%
\begin{equation}
\mathrm{curl}T^{ab}=\varepsilon ^{cd<a}D_{c}\sigma _{~~~d}^{b>}~.
\end{equation}

We expand all tensor quantities in terms of tensorial three-dimensional
harmonics with even $Q_{ab}^{\left( k\right) }$ and odd $\overline{Q}%
_{ab}^{\left( k\right) }$ parities. These harmonics are related by a $\mathrm%
{curl}$ operation [see Equations (A19) and (A20) of~\cite{Challinor2},
respectively, or Subsection 11.2.3 of \cite{EMM}] 
\begin{equation}
\mathrm{curl}Q_{ab}^{\left( k\right) }=\frac{k}{a}\sqrt{1+\frac{3K}{k^{2}}}%
\overline{Q}_{ab}^{\left( k\right) }~,  \label{curlQ}
\end{equation}%
\begin{equation}
\mathrm{curl}\overline{Q}_{ab}^{\left( k\right) }=\frac{k}{a}\sqrt{1+\frac{3K%
}{k^{2}}}Q_{ab}^{\left( k\right) }~.  \label{curlQoverb}
\end{equation}

The tensor harmonics are symmetric, trace-free and obey%
\begin{eqnarray}
D^{2}Q_{ab}^{\left( k\right) } &=&-\frac{k^{2}}{a^{2}}Q_{ab}^{\left(
k\right) }~,  \notag \\
D^{a}Q_{ab}^{\left( k\right) } &=&0~,~\dot{Q}_{ab}^{\left( k\right) }=0~,
\end{eqnarray}%
and the same relations hold for $\overline{Q}_{ab}^{\left( k\right) }~$[see
Appendix E and F of \cite{Ullrich}]. Here, $k$ is discrete for closed
Friedmann spacetimes while continous for flat and open universes \cite%
{Challinor2}. From Equations (\ref{curlQ}) and~(\ref{curlQoverb}) it follows
that $Q_{ab}^{\left( k\right) }$ satisfies the relation

\begin{equation}
\mathrm{curl}\,\mathrm{curl}Q_{ab}^{\left( k\right) }=\frac{k^{2}}{a^{2}}%
\left( 1+\frac{3K}{k^{2}}\right) Q_{ab}^{\left( k\right) }~,
\end{equation}%
and the same is true for $\overline{Q}_{ab}^{\left( k\right) }$.

All symmetric, trace-free and divergenceless first- order tensors are
expanded as%
\begin{equation}
T_{ab}=\sum\limits_{k}\left( T_{k}Q_{ab}^{\left( k\right) }+\overline{T}_{k}%
\overline{Q}_{ab}^{\left( k\right) }\right) ~.  \label{harmexp}
\end{equation}

Note that sometimes some power of $k/a$ are included in the harmonic
expansion \cite{Challinor2,perturb4}, and the~convention used in Equation (%
\ref{harmexp}) corresponds to that of \cite{DunsbyBassettEllis}. Since the
three-dimensional volume element occurs in the definition of $H_{ab}$, $%
\overline{H}_{k}$ belongs to the even parity sector, and $H_{k}$ to the odd
parity~sector.

By using the harmonic expansion the constraints (\ref{divfreeH}) and (\ref%
{divfreeE}) are satisfied, while~Equation~(\ref{Heq})~becomes%
\begin{eqnarray}
H_{k} &=&\frac{k}{a}\sqrt{1+\frac{3K}{k^{2}}}\overline{\sigma}_{k}~, \\
\overline{H}_{k} &=&\frac{k}{a}\sqrt{1+\frac{3K}{k^{2}}}\sigma _{k}~.
\end{eqnarray}

Applying these, the evolution equations Equations (\ref{sigmaEq})--(\ref%
{HabEq}) reduce to%
\begin{equation}  \label{dotsigma3dim}
\dot{\sigma}_{k}+\frac{2}{3}\Theta \sigma _{k}+E_{k}=0~,
\end{equation}%
\begin{equation}  \label{dotE3dim}
\dot{E}_{k}+\Theta E_{k}+\left[ \frac{\mu +p}{2}-\frac{k^{2}}{a^{2}}\left( 1+%
\frac{3K}{k^{2}}\right) \right] \sigma _{k}=0~,
\end{equation}%
\begin{equation}  \label{dotH3dim}
\dot{H}_{k}+\Theta H_{k}+\frac{k}{a}\sqrt{1+\frac{3K}{k^{2}}}\overline{E}%
_{k}=0~.
\end{equation}

Equations for the coefficients belonging to the odd parity can be obtained
by interchanging the overbared and unoverbared variables, respectively. From
Equations (\ref{dotsigma3dim})--(\ref{dotH3dim}), the following wave
equations can be derived for the shear, and the electric and magnetic parts
of Weyl tensor:%
\begin{equation}
\ddot{\sigma}_{k}+\frac{5}{3}\Theta \dot{\sigma}_{k}+\left[ \frac{k^{2}}{%
a^{2}}\left( 1+\frac{K}{2k^{2}}\right) +\frac{\Theta ^{2}}{6}+\frac{3}{2}%
\left( \Lambda -p\right) \right] \sigma _{k}=0~,  \label{s1}
\end{equation}%
\begin{equation}
\ddot{E}_{k}+q_{1}\dot{E}_{k}+q_{0}E_{k}=0~,  \label{E2}
\end{equation}%
\begin{equation}
\ddot{H}_{k}+\frac{7\Theta }{3}\dot{H}_{k}+\left[ \frac{k^{2}}{a^{2}}+\frac{%
2\Theta ^{2}}{3}+2\left( \Lambda -p\right) \right] H_{k}=0~,  \label{H3}
\end{equation}%
with coefficients%
\begin{equation}
q_{0}=\frac{k^{2}}{a^{2}}+\frac{2\Theta ^{2}}{3}+2\left( \Lambda -p\right) -%
\frac{\left( 1+3c_{s}^{2}\right) \left( \mu +p\right) }{\frac{2k^{2}}{a^{2}}%
\left( 1+\frac{3K}{k^{2}}\right) -\left( \mu +p\right) }\frac{\Theta ^{2}}{3}%
~,
\end{equation}%
\begin{equation}
q_{1}=\frac{7\Theta }{3}-\frac{\left( 1+3c_{s}^{2}\right) \left( \mu
+p\right) }{\frac{2k^{2}}{a^{2}}\left( 1+\frac{3K}{k^{2}}\right) -\left( \mu
+p\right) }\frac{\Theta }{3}~.
\end{equation}

Similar equations are valid for the other parity variables. In the
derivation, the time derivatives of $\mu $ and $\Theta $ were eliminated by
using Equations (\ref{drho}) and (\ref{dTheta}), respectively, the time
derivate of $p$ by%
\begin{equation}
\dot{p}=c_{s}^{2}\dot{\mu}~,
\end{equation}%
and finally the energy density $\mu $ is partly elimated by using (\ref%
{Theta}). Equations (\ref{s1})--(\ref{H3}) correspond to Equations (\ref%
{SigmaEqF}), (\ref{EEqF}) and (\ref{HF}), respectively, for $K=0$ and for
the vanishing energy density~2-gradient. 


\end{document}